\documentclass[sigplan,screen]{acmart}

\usepackage{my-package}

\setcopyright{cc}
\setcctype{by}
\acmDOI{10.1145/3779031.3779096}
\acmYear{2026}
\copyrightyear{2026}
\acmISBN{979-8-4007-2341-4/2026/01}
\acmConference[CPP '26]{Proceedings of the 15th ACM SIGPLAN International Conference on Certified Programs and Proofs}{January 12--13, 2026}{Rennes, France}
\acmBooktitle{Proceedings of the 15th ACM SIGPLAN International Conference on Certified Programs and Proofs (CPP '26), January 12--13, 2026, Rennes, France}
\received{2025-09-03}
\received[accepted]{2025-11-13}

\begin{document}

\newlength{\oldJot}
\setlength{\oldJot}{\jot}
\setlength{\jot}{1.5pt}

\title{Precise Reasoning about Container-Internal Pointers with Logical Pinning}

\author{Yawen Guan}
\orcid{0009-0007-5102-1724}
\affiliation{%
  \institution{EPFL}
  \city{Lausanne}
  \country{Switzerland}
}
\email{yawen.guan@epfl.ch}

\author{Clément Pit-Claudel}
\orcid{0000-0002-1900-3901}
\affiliation{%
  \institution{EPFL}
  \city{Lausanne}
  \country{Switzerland}
}
\email{clement.pit-claudel@epfl.ch}


\begin{abstract}
  Most separation logics hide container-internal pointers for modularity. This makes it difficult to specify container APIs that temporarily expose those pointers to the outside, and to verify programs that use these APIs.

  We present \emph{logical pinning}, a lightweight borrowing model for sequential programs that allows users to selectively track container-internal pointers at the logical level.  Our model generalizes the magic-wand operator for representing partial data structures, making it easy to write and prove precise specifications, including pointer-stability properties.  Because it only changes the way representation predicates and specifications are written, our approach is compatible with most separation logic variants.

  We demonstrate the practicality of logical pinning by verifying small but representative pointer-manipulating programs, and deriving more precise versions of common container specifications.  In doing so, we show that our approach subsumes some well-known proof patterns, simplifies some complex proofs, and enables reasoning about program patterns not supported by traditional specifications.  All of our results are mechanized in the Rocq proof assistant, using the CFML library.
\end{abstract}



\maketitle

\section{Introduction}\label{sec:introduction}

Separation logic~\cite{ohearnLocalReasoningPrograms2001c, reynoldsSeparationLogicLogic2002, ohearnSeparationLogic2019, charguéraudModernEyeSeparation2023} is an extension of Hoare logic for modular verification of pointer-manipulating programs. A central challenge in reasoning about such programs is aliasing, since mutating a shared memory cell may affect the value of syntactically unrelated expressions.

Separation logic helps avoid this challenge by enabling assertions over disjoint heap fragments using the separating conjunction operator~\(\star\), such that mutating a single memory cell in the heap affects only one \(\star\)-conjunct. The primitive heap proposition in the logic is \(\pointsto{p}{v}\), asserting exclusive \textit{ownership} (permission to access and mutate) of the cell at location \(p\) storing data \(v\). A \(\star\)-conjunction like \(\pointsto{p}{v} \starsp \pointsto{q}{u}\) ensures that \(p\) and \(q\) are not aliases: the \(\star\) captures the absence of sharing, enforcing non-aliasing by construction.

In separation logic, mutable containers such as linked lists and trees are described using predicates, called \textit{representation predicates}, which relate the container's logical model to a concrete heap representation. For modularity, most of them hide the container's internal pointers and assert full ownership of the container and its content.

\note{[space] can save space here}
For example, consider a linked list whose nodes are represented in C as:
\begin{lstlisting}[style=simple-ccode]
struct lnode { elem x; struct lnode* next; };
\end{lstlisting}
Assume that the C type \texttt{elem} corresponds to the logical type \(A\), and that the representation predicate \(\appTwo{R_A}{x}{p}\) asserts that an element \(x\) of type \(A\) is stored at \(p\). The typical representation predicate of such lists, \(\reprList{}\), can be defined as:
\begin{flalign*}
& \ReprList{R_A}{\nil}{p} \eqdef \hpure{p = \nullptr} \\
& \ReprList{R_A}{(\Cons{x}{l})}{p} \eqdef \appTwo{R_A}{x}{p} \starsp \hexistsx{q} \pointsto{p+1}{q} \starsp \ReprList{R_A}{l}{q}
\end{flalign*}
This asserts that the pointer \(p\) is \(\nullptr{}\) when the list is empty, and otherwise the head \(x\) is stored at location \(p\) and a pointer \(q\) pointing to the tail \(l\) is stored at \(p+1\). The recursive use of \(\star\) enforces that all list nodes occupy disjoint heap fragments. Such representations are convenient and effective when the non-aliasing assumption holds, namely, that each element is accessible only through a unique path rooted at the container's entry point.

Unfortunately, the non-aliasing assumption often breaks down in practice: container interfaces and implementations commonly expose internal pointers to elements (stored items) or substructures to the outside. Graphs are inherently built on sharing, whereas other containers expose internal pointers only temporarily, either to client code or during internal operations such as tree rebalancing. In this work, we focus on containers with transient internal-pointer exposure, both flat (e.g., arrays and records) and recursive (e.g., linked lists and trees), and develop techniques for reasoning about them in sequential programs written in pointer-manipulating languages such as C.

To illustrate the challenges concretely, consider the following logger example, inspired by the APIs of the C++ logging library \texttt{spdlog}~\cite{SpdlogFastLogging2025} and rewritten in C for simplicity. A logger is an object that maintains a configuration, such as log levels, and directs messages to designated outputs. Applications often manage a dictionary of named loggers; in this example, \texttt{registry} is a pointer to such a dictionary, whose loggers can be updated individually or collectively.
\begin{lstlisting}[style=ccode]
struct logger* lgr = get(registry, k);
// spdlog::set_level
set_level_all(registry, INFO);
// spdlog::logger::set_level
set_level(lgr, DEBUG);
\end{lstlisting}
Line 1 retrieves a pointer to the logger associated with key \texttt{k} and stores it in variable \texttt{lgr}; line 3 sets the log level of \textit{all} loggers in \texttt{registry} to \texttt{INFO}; and line 5 overrides the level of the logger pointed to by \texttt{lgr} to \texttt{DEBUG}. This program is safe because the implementation of \setlevelall{} ensures pointer stability: it updates loggers in place, so the pointer \texttt{lgr} remains valid for use on line 5.

Conceptually, \setlevelall{} is a map operation that updates all loggers in the dictionary. Let \(\ReprDict{d}{p}\) denote the representation predicate for a dictionary \(d\) stored at \(p\), and \(\loggerSubstLevel{g}{v}\) denote the logger \(g\) with its level substituted by \(v\). A typical specification of \setlevelall{} is:
\begin{flalign*}
& \forall~d~p~v.\; \tstate{\ReprDict{d}{p}}\; \setLevelAll{p}{v} \; \\
&&& \mathllap{
    \tstateUnit{\ReprDict{(\Mapf{g}{\loggerSubstLevel{g}{v}}{d})}{p}}
    }
\end{flalign*}
Unfortunately, this specification is insufficient to verify the program: it does not capture pointer stability, and hence permits implementations that relocate (move) loggers and invalidate exposed internal pointers such as \texttt{lgr}.


Existing approaches fall short on verifying this pattern for various reasons (\cref{sec:background}). In this paper, we introduce a \textit{logical-pinning} model for reasoning about exposed internal pointers, together with a \textit{proof discipline} in which the user selectively exposes such pointers in the logical representations and precisely tracks pointer validity in API specifications.


\Cref{sec:overview} outlines the key ideas of our approach. \Cref{sec:background} discusses the limitations of existing state-of-the-art approaches. \Cref{sec:logical-pinning-model} presents the logical-pinning model and proof discipline. \Cref{sec:case-studies,sec:discussion} apply the model to a series of containers and discuss the results, showing that it subsumes existing ad hoc proof patterns, allows complex proofs to be simplified, and enables verification of new program patterns. We conclude with related work in \cref{sec:related-work}.

\subsubsection*{Contributions}
\begin{itemize}
  \item We introduce \textit{logical pinning}, a model and proof discipline for precise reasoning about exposed internal pointers in containers.
  \item We showcase applications of the model in a series of case studies, covering containers ranging from single cells to arrays, records, linked lists, and trees.
  \item We show that logical-pinning enables maximally precise specifications for realistic APIs by specifying a realistic, low-level API for linked lists.
  \item We show that logical pinning is easy to implement in practice on top of existing separation logics by formalizing all of our examples using the CFML library.
\end{itemize}

\todo{realistic list API spec: is it the first time?}

\section{Overview and Key Ideas}\label{sec:overview}

\begin{figure}[htbp]
  \centering
  \includegraphics[
    width=0.8\linewidth,
    trim=20 5 20 10, 
    clip
  ]{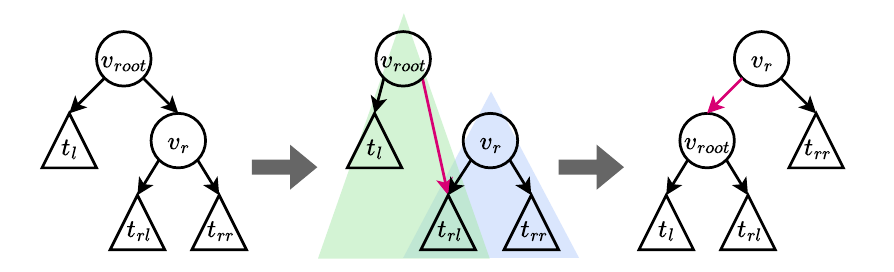}
  \caption{Left rotation of a binary tree.}\label{fig:tree-left-rotation}
  \Description{Left rotation of a binary tree.}
\end{figure}

Many programming idioms require containers to temporarily expose their internal pointers to the outside:
\begin{itemize}
  \item Pointer caching, to avoid repeated lookups in performance-critical code;
  \item Direct manipulation, such as swapping two values stored in a list after retrieving pointers to them;
  \item Transient cross-container sharing, such as binary-tree rotation (\cref{fig:tree-left-rotation}) where a subtree \(t_{rl}\) is temporarily shared between two trees.
\end{itemize}

Exposing internal pointers raises two challenges: (1) reasoning about their validity under container operations, and (2) supporting \textit{multiple borrows}, where ownership of several subparts referenced by exposed pointers may be temporarily detached from the container for local reasoning. Existing approaches, such as magic wands, fall short on resolving these challenges and are discussed further in \cref{sec:background}.

\subsection{Key Idea: Capture Pointer Stability with a Precise Classification of Writes}\label{key-idea:classify-writes}

\textit{Pointer stability} is a property of a container operation that ensures pointers to its subparts remain valid, i.e., the subparts are neither deallocated nor relocated. This property allows clients to safely cache and reuse pointers across operations.

\begin{figure}[htbp]
  \centering
  \makebox[\linewidth][c]{
    \includegraphics[
    width=\linewidth,
    trim=50 5 30 10, 
    clip
    ]{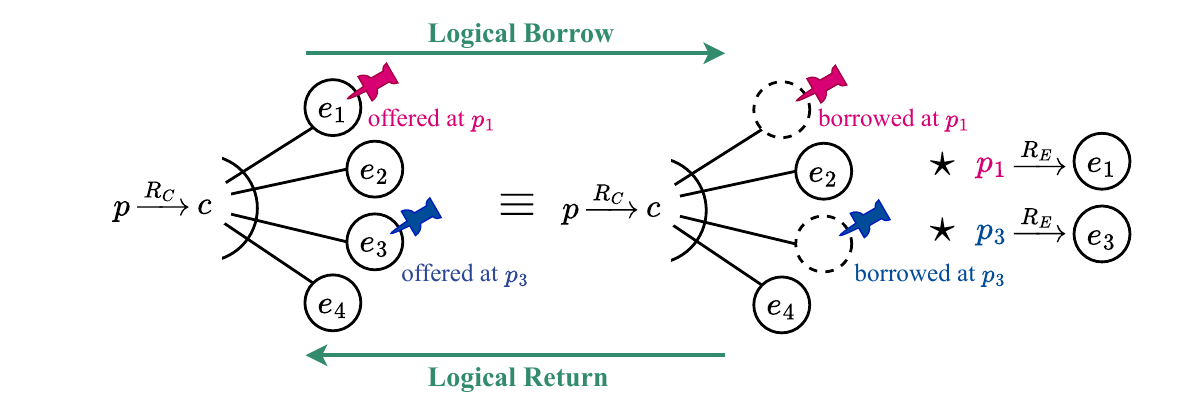}
  }
  \caption{Two views of a container \(c\) with elements \(e_1 \dots e_4\) and exposed internal pointers \(p_1\) and \(p_3\). Left: \(c\) owns all elements. Right: \(e_1\) and \(e_3\) are borrowed from \(c\). }\label{fig:offered-borrowed}
  \Description{Two views of a container, linked by logical borrow and return.}
\end{figure}

The traditional read-write model, which broadly classifies memory operations as reads and writes, is too coarse to reason about pointer stability because it fails to distinguish between fundamentally different types of writes:
\begin{itemize}
  \item \textbf{\textit{In-place writes}}: Writes that perform in-place mutation, such as the \setlevelall{} operation. They change the content without invalidating existing pointers.
  \item \textbf{\textit{Full writes}}: Writes that may deallocate or relocate objects, such as resizing, rebalancing, or restructuring containers. They can invalidate previously exposed pointers.
\end{itemize}
Many container APIs (e.g., in C++ STL) explicitly document whether an operation invalidates internal pointers~\cite{ContainersLibraryCppreferencecom2025}. Correspondingly, we would like to capture in specifications which kind of writes a function may perform. Consider a container \(c\) consisting of four elements \(e_1 \dots e_4\), where only two internal pointers are exposed: \(p_1\) to \(e_1\) and \(p_3\) to \(e_3\) (left of \cref{fig:offered-borrowed}; the right is discussed in \cref{model:borrowing-state-transitions}). For readability, we write \(\repr{R}{x}{p}\) for application \(\R{x}{p}\) in figures. Conceptually, any function \(f\) that manipulates \(c\) while promising to preserve these exposed pointers must treat \(e_1\) and \(e_3\) as \textit{pinned} at their location, meaning \(f\) may only perform reads and in-place writes to them. In contrast, \(e_2\) and \(e_4\) are considered \textit{floating}; \(f\) may perform full writes to them.

\begin{figure}[htbp]
  \centering
  \makebox[\linewidth][c]{
    \includegraphics[
    width=\linewidth,
    trim=0 0 0 0, 
    clip
    ]{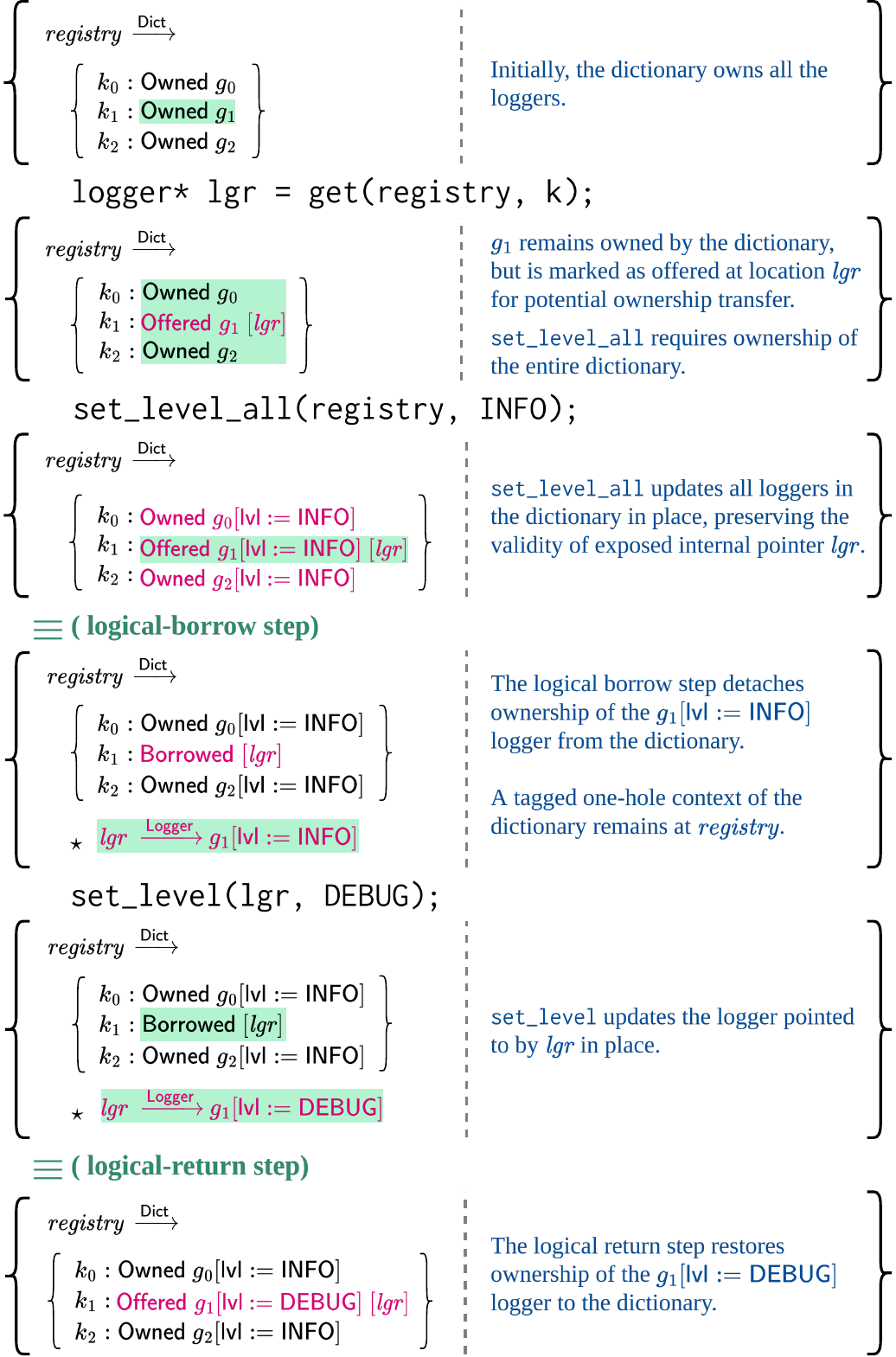}
  }
  \caption{An informal proof sketch of the functional correctness of the logger example. For simplicity, we assume the dictionary initially contains three key-logger pairs \(\tupleTwo{k_0}{g_0}\), \(\tupleTwo{k_1}{g_1}\) and  \(\tupleTwo{k_2}{g_2}\), where \(k = k_1\). Each step is annotated with the corresponding abstract heap state, denoted \(\tstate{\textit{state}}\), together with a brief informal explanation (shown in blue). \([\lgr]\) denotes an alias list with a single element \(\lgr\). Color markers: \tochangetext{to change}, \changed{just changed}.}\label{fig:logger-informal-proof}
  \Description{An informal proof sketch of the functional correctness of the logger example.}
\end{figure}

\subsection{Key Idea: Encode Representation Predicates as Tagged Multi-Hole Contexts}\label{key-idea:multi-hole-context}

In practice, programs often maintain multiple exposed pointers to container subparts (elements and substructures) rather than just one. A simple example is the \textit{element-swapping} program which retrieves pointers to two list elements and swaps their contents. When ownership of these subparts is detached for local reasoning (i.e., these subparts are \textit{borrowed}), the container is left with multiple holes.

Consider a key-value tree \(t\) pointed to by \(p\). The program below looks up potentially overlapping subtrees \(t_1\) and \(t_2\) rooted at keys \(k_1\) and \(k_2\) before passing them on to \texttt{compare}:
\begin{lstlisting}[style=ccode]
struct tree* q1 = lookup(p, k1);
struct tree* q2 = lookup(p, k2);
compare(q1, q2);
\end{lstlisting}

A typical specification for \(\codeAppTwo{\texttt{lookup}}{p}{k_1}\) requires the full ownership of \(t\), and splits that ownership between \(t_1\) (rooted at \(q_1\)) and a partial tree rooted at \(p\).  This partial tree is often represented using a custom predicate or a magic wand.

This specification performs two logical steps at once: it records the subtree relation between \(q_1\) and \(p\), and transfers ownership of \(t_1\) to \(q_1\).
This ownership transfer is incompatible with the second lookup, which requires \(p\) to have full ownership of \(t\).

Our approach distinguishes these two steps: in it, the calls to \texttt{lookup} pin \(t_1\) to \(q_1\) and \(t_2\) to \(q_2\) without detaching them, and without changing the predicate used to represent \(t\).
Later, when separate ownership is needed to call \texttt{compare}, we add a \textit{logical borrowing} step to detach \(t_1\) and \(t_2\).

To do so, we adopt a uniform representation for whole and partial containers using \textit{tagged multi-hole contexts}, in which each borrowable subpart is augmented with a fine-grained \textit{borrowing state} that indicates whether the subpart is pinned or floating, and whether the container still has its ownership.  This allows precise specifications that support flexible pointer tracking and ownership transfers for individual subparts without breaking the container abstraction.

This subtree-compare example is verified using our approach; the key to handling potential overlapping subtrees is the use of non-separating conjunction (\cref{bg:separation-logic}) in the precondition of \texttt{compare}. See \cref{app:subtree-compare} for details.

\subsection{The Logger Example Revisited}\label{sec:revisit}

\Cref{fig:logger-informal-proof} presents an informal correctness proof of the logger example under our approach. The predicate \(\Logger{g}{p}\) asserts that the logger \(g\) is stored at \(p\). We write specifications with implicit universal quantification over free variables (e.g., \(g\), \(p\), \(v\) below). The specification of \setlevel{} is:
\[
\tstate{\Logger{g}{p}}\, \setLevel{p}{v}\, \tstateUnit{ \Logger{\loggerSubstLevel{g}{v}}{p}}
\]
We tag each logger in the dictionary with one of the four borrowing states: \(\owned\), \(\offered\), \(\borrowed\) or \(\missing\). We refine the specification of \texttt{get} to mark the returned pointer \(\lgr\) by offering the owned logger \(g_1\) at \(\lgr\), essentially pinning it without detaching its ownership from the dictionary. We update the specification of \setlevelall{} to preserve all pinned locations in the dictionary, ensuring the validity of exposed internal pointers.

\section{Background}\label{sec:background}

In this section, we first briefly introduce separation logic, then review and discuss the limitation of the main state-of-the-art approaches for reasoning about access to and mutation of container subparts.

\subsection{Separation Logic}\label{bg:separation-logic}

Separation logic is a refinement of Hoare logic that enables local reasoning: one can prove a Hoare triple by describing only the relevant heap fragment and apply the triple in a larger context. We use a standard shallow embedding of a linear separation logic in a higher-order logic, which models heaps (heap fragments) as partial functions from locations to values, with type \(\loc{} \rightarrow \val{}\) (abbreviated as \(\heap{}\)).

We assume the existence of a distinguished \(\nullptr\) pointer and allow pointer arithmetic on locations, so an expression like \(p+n\) denotes the location \(n\) steps offset from the base location \(p\).
We denote the empty heap as \(\emptyset\). Two heaps \(h_1\) and \(h_2\) are disjoint (written \(h_1 \bot h_2\)) when their domains do not overlap, and can be combined by union, denoted as \(h_1 \uplus h_2\).

{
\setlength{\abovecaptionskip}{5pt}
\setlength{\belowcaptionskip}{-2pt}
\begin{figure}[htbp]
\begin{flalign*}
  \hpure{P} &\eqdef \lambdax{h} h = \emptyset \andsp P  \\
  \pointsto{p}{v} &\eqdef \lambdax{h} h = (p \rightarrow v) \andsp p \neq \nullptr \\
  H_1 \star H_2 &\eqdef \lambdax{h}
                  h_1 ~\bot~ h_2 \andsp
                  h_1 \uplus h_2 \andsp
                  \appOne{H_1}{h_1} \andsp
                  \appOne{H_2}{h_2} \\
  \hexistsx{x} H &\eqdef \lambdax{h} \existsx{x} \appOne{H}{h} \qquad
  \hforallx{x} H \eqdef \lambdax{h} \forallx{x} \appOne{H}{h} \\
  H_1 \wand H_2 &\eqdef \hexistsx{H} H \star \hpure{(H_1 \star H) \vdash H_2} \\
  & \qquad \text{where \(H_1 \vdash H_2 \eqdef \forallx{h} \appOne{H_1}{h} \rightarrow \appOne{H_2}{h}\)} \\
  H_1 \hand H_2 &\eqdef \lambdax{h} \appOne{H_1}{h} \wedge \appOne{H_2}{h} \quad
  H_1 \hor H_2 \eqdef \lambdax{h} \appOne{H_1}{h} \vee \appOne{H_2}{h}
\end{flalign*}
\caption{Basic heap predicate combinators.}\label{fig:heap-predicate-combinators}
\Description{Basic heap predicate combinators.}
\end{figure}
}

A heap predicate \(H\) is a function of type \(\heap{} \rightarrow \prop{}\) (abbreviated as \(\hprop{}\)), used to describe properties of heaps. \Cref{fig:heap-predicate-combinators} shows the basic heap predicate combinators.
\(\hpure{P}\) asserts a pure (non-spatial) proposition \(P\). \(\pointsto{p}{v}\) denotes a singleton heap mapping location \(p\) to value \(v\). The separating conjunction \(H_1 \star H_2\) holds when the heap can be divided into two disjoint parts satisfying \(H_1\) and \(H_2\) respectively. \(\hexistsx{x}{H}\) and \(\hforallx{x}{H}\) are existential and universal quantification over values in the heap (\(\star\) has higher precedence than \(\hexists\) and \(\hforall\)). The magic wand (separating implication), \(H_1 \wand H_2\), denotes a heap that, when extended with any heap satisfying \(H_1\), yields a heap satisfying \(H_2\). The non-separating conjunction \(H_1 \hand H_2\) asserts that the heap satisfies \(H_1\) and \(H_2\), while the disjunction \(H_1 \hor H_2\) asserts it satisfies at least one of them.

We specify programs using Hoare triples of the form \(\triplex{e}{H}{v}{\appOne{Q}{v}}\) where \(e\) is an expression, \(H\) is the precondition describing the heap before execution, \(v\) is the return value of \(e\), and \(\appOne{Q}{v}\) is the postcondition describing the heap after the execution of \(e\). This triple asserts total correctness: if the initial heap satisfies \(H\), then \(e\) will terminate and produce an output value \(v\) and a heap fragment satisfying \(\appOne{Q}{v}\).

\todo{Move total correctness to implementation?}

A representation predicate \(R_A\) relates the logical model of an object with a location (its entry point) and a heap fragment (the memory it occupies). Formally, \(R_A\) has type \(A \rightarrow \loc \rightarrow \hprop\), abbreviated as \(\ReprType{A}\).

\subsection{Magic Wands}\label{bg:magic-wand}

In separation logic, detaching ownership (borrowing) is traditionally modeled using a magic wand \(\wand\). This operator is well-suited for a single immutable borrow, where the borrowed subpart can later be returned unchanged.

In the logger example, a possible but unhelpful specification of \texttt{get} tracks the returned pointer using a magic wand:
\begin{flalign*}
  & \tstate{\hpure{d[k] = g} \star \ReprDict{d}{p}}\; \codeAppTwo{\texttt{get}}{p}{k} \\
  &&& \mathllap{
        \tstate{\lambdax{r} \Logger{g}{r} \starsp (\Logger{g}{r} \wand \ReprDict{d}{p})}
      }
\end{flalign*}
\(d[k]\) denotes the logger associated with key \(k\) stored in dictionary \(d\). This specification is limited: after performing \texttt{get}, operations that require the dictionary representation \(\ReprDict{d}{p}\), such as \texttt{get} and \setlevelall{}, cannot be performed until the current wand is canceled. However, wand cancelation logically forgets the exposed pointer and thus invalidates it. Moreover, if the borrowed logger is mutated, the wand cannot be canceled at all, leaving the dictionary inaccessible.
\[
  \textit{\small (wand cancellation)}\quad H_1 \star (H_1 \wand H_2) \vdash H_2
\]
Therefore, when verifying the logger example with this specification, \(\lgr\) is invalid after calling \setlevelall{}, causing the verification to get stuck at \texttt{set\_level(lgr, DEBUG)}.

\paragraph{Magic wand as frame}\label{bg:magic-wand-as-frame}

The \textit{magic-wand-as-frame}~\cite{caoProofPearlMagic2019, wangCertifyingGraphmanipulatingPrograms2019a} technique extends the traditional use of magic wand to support a single mutable borrow. Instead of fixing the borrowed value, it universally quantifies over all possible updates, allowing the value to be modified and later reintegrated.

Under this approach, \texttt{get} can be specified as:
\begin{flalign*}
  & \tstate{ \hpure{d[k] = g} \star \ReprDict{d}{p}}\; \codeAppTwo{\texttt{get}}{p}{k}\; \\
  &&& \mathllap{
        \tstatex{r}{\Logger{g}{r} \star (\forallx{g'} \Logger{g'}{r} \wand \ReprDict{\subst{d}{k}{g'}}{p})}
      }
\end{flalign*}
\(\subst{d}{k}{g'}\) denotes the dictionary \(d\) with the logger associated with key \(k\) substituted by \(g'\).

This specification allows the borrowed logger \(g\) to be updated in place and returned to the dictionary afterward. However, the magic-wand-as-frame technique still does not support multiple borrows: a second call to \texttt{get} again requires canceling the wand and thereby revoking the first borrow.

\subsection{Trees with Holes and Cut Subtrees}\label{bg:trees-with-holes-and-cuts}

\todo{discuss nested borrow: holes and cuts does not work when the path are invalidated}

\todo{Critical difference: model for ownership transfer of subparts, but single step}

Charguéraud~\cite{charguéraudHigherorderRepresentationPredicates2016} introduces a technique to take multiple mutable borrows of elements or subtrees within a tree. Consider a binary tree whose nodes are represented in C as:
\begin{lstlisting}[style=simple-ccode]
struct tnode { elem x; struct tnode *lt, *rt; };
\end{lstlisting}
In addition to the standard constructors for leaves and nodes, datatype \(\holeCutTree\) includes two additional constructors:
\begin{flalign*}
  \HoleCutTree{A} \;\eqdef\; &\leaf \;|\; \appTwo{\node}{(x:A)}{(t_l, t_r: \HoleCutTree{A})} \\
  |\; &\appTwo{\hole}{(q:\loc)}{(t_l, t_r: \HoleCutTree{A})} \;|\; \Cut{(q: \loc)}
\end{flalign*}
\(\Node{x}{t_l}{t_r}\) denotes a node with element \(x\), left and right subtrees \(t_l\) and \(t_r\); \(\Hole{q}{t_l}{t_r}\) denotes a node whose element stored at \(q\) is absent, while the subtrees remain owned; and \(\cut\) denotes an absent subtree at \(q\).

The type \(\holeCutTree\) models trees with possibly absent elements (holes) and subtrees (cuts), and thus provides a uniform representation for whole and partial trees. Each hole and cut is annotated with its location to enable reattachment. The tree representation is the same as for ordinary trees, except that if a tree stored at \(p\) is \(\Hole{q}{t_l}{t_r}\) or \(\Cut{q}\), then \(p = q\) and the ownership of the absent element or subtree is not held.

This representation models ownership transfer in a single step: an element or subtree is either present or replaced by a hole or a cut. However, this granularity is too coarse for patterns such as the subtree–compare example in \cref{key-idea:multi-hole-context}. There, specifying \(\codeAppTwo{\texttt{lookup}}{p}{k_1}\) with this representation requires converting \(t_1\) into a cut to track its pointer \(q_1\), thereby detaching its ownership from \(t\). This detachment is undesirable, since it either (1) blocks a second \texttt{lookup}, or (2) forces the restoration of \(t_1\) into \(t\), which hides the location of \(t_1\) behind the representation of \(t\) and thus logically forgets and invalidates \(q_1\) before it is used in \texttt{compare}.



\subsection{Borrows in Rust}\label{bg:rust-borrows}

The pattern of exposing and manipulating internal pointers of containers is common in low-level programming. The same pattern naturally arises in Rust, where the type system enforces memory safety through controlled aliasing.

We return in \cref{sec:related-work} to Rust’s borrowing model, including the standard borrows, splitting borrows and deferred borrows. While these mechanisms provide strong safety guarantees, they remain insufficient to express certain patterns of internal pointer exposure: current Rust cannot easily express the above examples (logger updates, element swaps, or borrowing multiple elements and subtrees in a tree). The Rustonomicon calls our element-swapping example “pretty clearly hopeless” for “general container types like a tree”~\cite{therustprojectdevelopersSplittingBorrowsRustonomicon2025}.

\todo{Can deferred borrow express the logger example? Check it out.}

\section{The Logical-Pinning Model for Reasoning About Internal Pointers}\label{sec:logical-pinning-model}

In this section, we present \emph{logical pinning}, a lightweight borrowing model for reasoning about sequential programs that expose and manipulate pointers to container subparts (i.e., elements and substructures). At its core, the model defines \textit{fine-grained borrowing states} along with transitions that describe how these states evolve in response to program operations or logical reasoning steps.

Building on this model, we introduce a proof discipline in which users (1) identify borrowable subparts and redefine the container's representation predicate to track their borrowing states; (2) enrich API specifications to reflect the validity of exposed internal pointers; and (3) prove correctness of API and client programs with explicit logical borrow and return steps around subpart access. Our technique is formulated entirely within the basic form of separation logic defined in \cref{bg:separation-logic}, and does not require language-level extensions.

\subsection{Informal Overview}\label{model:informal-overview}

\todo{More vertical space after table.}

Each borrowable subpart has two orthogonal properties: (1) whether this subpart is \textit{available} to the container owner, i.e., the container owner has ownership of this subpart and therefore can access its logical value \(v\); (2) whether it is \textit{pinned} at an exposed memory location \(q\). The combination of these properties yields four borrowing states for each borrowable element or substructure of a container:
\begin{itemize}
  \item \boldmath{\(\Owned{v}\)}\unboldmath: The subpart is available and unpinned. The container owner may freely relocate it without affecting its logical state.

  \item \boldmath{\(\Offered{v}{q}{qs}\)}\unboldmath: The subpart is available but pinned to a specific exposed location \(q\). The list \(qs\) (possibly empty) represents additional exposed aliasing pointers.

  In this state, the container owner retains ownership of this subpart, but exposes its location for others to borrow it in the future. To preserve the same exposed location, the owner may perform only in-place updates, not relocation.

  \item \boldmath{\(\Borrowed{q}{qs}\)}\unboldmath: The subpart is pinned to the exposed location \(q\) (with additional aliases \(qs\)) and currently borrowed by another party. The subpart is unavailable to the container owner until it is returned at the same location.

  \item \boldmath{\(\Missing\)}\unboldmath: The subpart is unavailable and not tracked.
\end{itemize}
The \(\offered\) and \(\borrowed\) states are particularly important for modeling ownership transfer. One can decompose a container \(c\) that holds offered elements as follows:
\begin{flalign*}
& \textit{\(\mathrm{c}\) with element \(\mathrm{e}\) offered at \(\mathrm{q}\) is stored in memory} \\
\equiv \quad & \textit{\(\mathrm{c}\) with an element borrowed at \(\mathrm{q}\) is stored in memory} \\
& \quad \star \textit{\(\mathrm{e}\) is stored at \(\mathrm{q}\)}
\end{flalign*}
The forward direction corresponds to a \textit{logical borrow}, and the backward direction to a \textit{logical return}.

We introduce the \(\borrowable\) datatype to lift a base logical type \(A\) to its borrow-aware form \(\BorrowableFull{A}\) (written \(\Borrowable{A}\)) that models the four borrowing states described above. We also define a predicate transformer \(\liftToB\) that lifts the corresponding base representation predicate \(R_A\) to \(\LiftToB{R_A}\) (written \(\yB{R_A}\)) that describes borrowable objects of type \(\Borrowable{A}\). Full definitions are provided later in \cref{model:borrowing-states}.
\centerline{\begin{tabular}{ccc}
\toprule
& type & representation predicate \\
\midrule
base object & \(A\) & \(R_A: \ReprType{A}\) \\
borrowable object & \makecell[c]{\(\Borrowable{A}\)} & \makecell[c]{\(\yB{R_A}: \ReprType{\Borrowable{A}}\)} \\
\bottomrule
\end{tabular}}

\subsection{Step-by-Step Guide}\label{model:step-by-step-guide}

We illustrate the use of our model and proof discipline step by step using the motivating logger example introduced in \cref{sec:introduction}. For simplicity, assume that logger keys are integers. Suppose that the dictionary is implemented as a linked list of key–logger pairs, whose nodes are represented in C as:
\begin{lstlisting}[style=simple-ccode]
struct dnode { int key; struct logger* log;
                        struct dnode*  nxt; };
\end{lstlisting}
A typical representation predicate for such dictionaries is:
\newcommand{\dictCons}[1]{(\Cons{\tupleTwo{k}{#1}}{d})}
\begin{flalign*}
  &\; \reprDict: \reprDictType{\loggerType} &&\\
  &\; \ReprDict{\nil}{p} \eqdef \hpure{p = \nullptr} &&\\
  &\; \ReprDict{\dictCons{g}}{p}
    \begin{aligned}[t]
    & \eqdef \pointsto{p}{k} \\
    & \starsp \hexistsx{q_1} \pointsto{p+1}{q_1} \starsp \Logger{g}{q_1} \\
    & \starsp \hexistsx{q_2} \pointsto{p+2}{q_2} \starsp \ReprDict{d}{q_2} \\
    \end{aligned} &&
\end{flalign*}
When the dictionary is non-empty, the head key \(k\) is stored at \(p\), a pointer to the head logger \(g\) is stored at \(p+1\), and a pointer to the rest of the dictionary \(d\) is stored at \(p+2\).

\paragraph{Step 0. Abstract indirection and pointer arithmetic}

The ``\fieldname{log}'' and ``\fieldname{nxt}'' fields store intermediate pointers rather than embedding the referenced objects in place. We express such indirection using the \textit{indirection transformer} ``\(\yI{}\)'' (\cref{model:indirection}), so the definition below desugars to the one above. Changes are marked using the ``\changed{just changed}'' color convention from \cref{fig:logger-informal-proof}.
\begin{flalign*}
  &\; \reprDict: \reprDictType{\loggerType} &&\\
  &\; \ReprDict{\nil}{p} \eqdef \hpure{p = \nullptr} &&\\
  &\; \ReprDict{\dictCons{g}}{p} \eqdef &&\\
  &&& \mathllap{\pointsto{p}{k} \star \changed{\iLogger{g}{(p+1)}} \star \changed{\iReprDict{d}{(p+2)}}}
\end{flalign*}

We also use records (\cref{case-studies:arrays-and-records}) to syntactically encapsulate pointer arithmetic, allowing \(\reprDict\) to be further rewritten as below. Here, \(\cell{}\) is the points-to predicate asserting a value occupies a single cell:
\[ \forall~v~p.\; \Cell{v}{p} \eqdef \pointsto{p}{v} \]
\begin{flalign*}
  &\; \reprDict: \reprDictType{\loggerType} &&\\
  &\; \ReprDict{[]}{p} \eqdef \hpure{p = \nullptr} &&\\
  &\; \ReprDict{\dictCons{g}}{p} \eqdef &&\\
  &&& \mathllap{\changed{\Record{\recordDict{\cell}{k}{\ilogger}{g}{\yI{\reprDict}}{d}}{p}}}
\end{flalign*}
When the dictionary is non-empty, \(p\) points to a record with three fields: the ``\fieldname{key}'' field stores \(k\) within a single memory cell described by \(\cell{}\); the ``\fieldname{log}'' and ``\fieldname{nxt}'' fields store \(g\) and \(d\) indirectly, that is, store an intermediate pointer to the corresponding objects.

\paragraph{Step 1. Identify borrowable subparts and redefine the container’s representation predicate to track their borrowing states}

We first identify the borrowable subparts, in this case the loggers, and lift their logical models (of type \(\loggerType\)) to the borrow-aware form (of type \(\yB{\loggerType}\)), where each logger is annotated with its borrowing state. We use \(\bvar{v}\), \(\bpred{R}\) and \(\bpred{f}\) to denote the borrow-aware forms of the logical value \(v\), predicate \(R\) and function \(f\), respectively. We then define a borrow-aware representation predicate \(\reprDictE\) to describe a dictionary whose elements are borrowable:
\newcommand{\content}{
  \recordDict{\cell}{k}
    {\yI{\yBchanged{\logger}}}{\bg}
    {\yI{\reprDictEChanged}}{d}
}
\begin{flalign*}
  &\; \reprDictEChanged: \reprDictType{\BorrowableChanged{\loggerType}} &&\\
  &\; \ReprDictEChanged{\nil}{p} \eqdef \hpure{p = \nullptr} &&\\
  &\; \ReprDictEChanged{\dictCons{\bg}}{p} \eqdef &&\\
  &&& \mathllap{\Record{\content}{p}}
\end{flalign*}
This definition states that when the dictionary is non-empty, the ``\(\fieldname{log}\)'' field stores a borrowable logger \(\bg{}\) indirectly.

\paragraph{Step 2. Enrich API specifications to reflect the validity of exposed internal pointers}

To capture the fact that \(\setlevelall{}\) updates loggers in place, we refine its specification to explicitly ensure pointer stability:
\newcommand{\dictAvailable}{\forallx{\bg \in d}\, \isAvailable{\bg{}}}
\newcommand{\dictMapD}[1]{\Mapf{\bg}{\bloggerSubstLevel{\bg}{#1}}{d}}
\begin{flalign*}
  & \tstate{\hpure{\dictAvailable} \star \ReprDictE{d}{p}}\; \\
  & \qquad \setLevelAll{p}{v}\; \\
& \tstateUnit{\ReprDictE{(\dictMapD{v})}{p}} \end{flalign*}
\(\isAvailable{\bg{}}\) asserts that the underlying value of \(\bg{}\) is accessible, i.e., \(\bg\) is in the \(\owned{}\) or \(\offered{}\) state. This specification requires the availability of all loggers in the dictionary, allowing them to be offered at exposed locations. In the postcondition, \(\bloggerSubstLevel{\bg}{v}\) denotes a borrow-aware substitution that preserves exposed locations.

We can now specify \texttt{get} to track the returned pointer in the borrowing state without detaching \(g\) (\(\ConsNil{r}\) denotes \(\Cons{r}{\nil}\)):
\begin{flalign*}
  & \textit{\small (weakest)}\;
    \tstate{ \hpure{d[k] = \Owned{g}} \star \ReprDictE{d}{p}}\; \\
  &&& \mathllap{
        \codeAppTwo{\texttt{get}}{p}{k}\;
        \tstatex{r}{\ReprDictE{d[k := \OfferedOne{g}{r}]}{p}}
      }
\end{flalign*}
This first specification allows loggers other than \(d[k]\) to be unavailable; but it prohibits repeated calls to \texttt{get} with the same key, as it requires \(d[k]\) to be in the \(\owned\) state. We can strengthen it to allow the queried logger to be either \(\owned\) or \(\offered\):
\begin{flalign*}
  & \textit{\small (weaker)}\;
    \tstate{ \hpure{\changed{d[k] = \bg{} \wedge \isAvailable{\bg{}}}} \star \ReprDictE{d}{p}}\; \\
  &&& \mathllap{
        \codeAppTwo{\texttt{get}}{p}{k}\;
        \tstatex{r}{\ReprDictE{d[k := \changed{\Pin{\bg{}}{r}}]}{p}}
      }
\end{flalign*}
Function \(\Pin{\bv{}}{r}\) pins a borrowable value \(\bv{}\) at \(r\). The predicate transformer ``\(\yB\)'' ensures that if \(\bv{}\) is already pinned, the existing and new locations coincide.
\begin{center}
  \begin{tabular}{ll}
  \toprule
  $\bv{}$ & $\Pin{\bv{}}{r}$ \\
  \midrule
  $\Owned{v}$           & $\Offered{v}{r}{\nil}$  \\
  $\Offered{v}{q}{qs}$      & $\Offered{v}{r}{\Cons{q}{qs}}$ \\
  $\Borrowed{q}{qs}$        & $\Borrowed{r}{\Cons{q}{qs}}$   \\
  $\missing$            & $\Borrowed{r}{\nil}$   \\
  \bottomrule
  \end{tabular}
\end{center}
We can further strengthen the specification by allowing the queried logger to be \(\borrowed\) or \(\missing\) as well:
\begin{flalign*}
  & \textit{\small (strongest)}\;
    \tstate{ \hpure{\changed{d[k] = \bg{}}} \star \ReprDictE{d}{p}}\; \\
  &&& \mathllap{
        \codeAppTwo{\texttt{get}}{p}{k}\;
        \tstatex{r}{\ReprDictE{d[k := \Pin{\bg{}}{r}]}{p}}
      }
\end{flalign*}

\paragraph{Step 3. Prove correctness with explicit logical borrow and return steps around subpart access}

\Cref{fig:logger-informal-proof} illustrates a correctness proof of the logger example, with the correctness theorem stated as:
\begin{flalign*}
  & (\dictAvailable) \wedge (\bvar{g_k} = d[k]) \rightarrow \\
  & \text{let } d_1 \eqdef \dictMapD{\texttt{INFO}} \text{ in} \\
  & \tstate{\ReprDictE{d}{\registry}}\; (\text{the logger program})\; \\
  &\quad \tstateUnit{\ReprDictE{d_1[k := \Pin{\bloggerSubstLevel{\bvar{g_k}}{\texttt{DEBUG}}}{\lgr}]}{\registry}}
\end{flalign*}

Immediately before \setlevel{}, we perform logical borrow to detach the target logger’s ownership from the dictionary. From the dictionary’s perspective, the logger transitions from being offered at \(\lgr\) (available for future borrowing) to being borrowed at \(\lgr\) (held by another party), with \(\lgr\) enabling future return.

After performing \setlevel{}, whose specification frames out the partial dictionary, we perform a logical-return step (using the handle \(\lgr\)) to restore the updated logger's ownership to the dictionary, thereby recovering the full dictionary representation for subsequent operations.

\subsection{Borrowability}

In flat containers such as arrays and records, borrowable subparts are typically the stored items (elements), such as record fields and array entries. They can be individually manipulated without affecting the overall structure. In recursive containers such as lists and trees, borrowable subparts include both elements \textit{and} recursively defined substructures, such as list tails and subtrees.

\subsubsection{Borrowing States}\label{model:borrowing-states}

\begin{figure}[htbp]
  \centering
  \makebox[\linewidth][c]{
    \includegraphics[
    width=\linewidth,
    trim=10 10 20 10, 
    clip
    ]{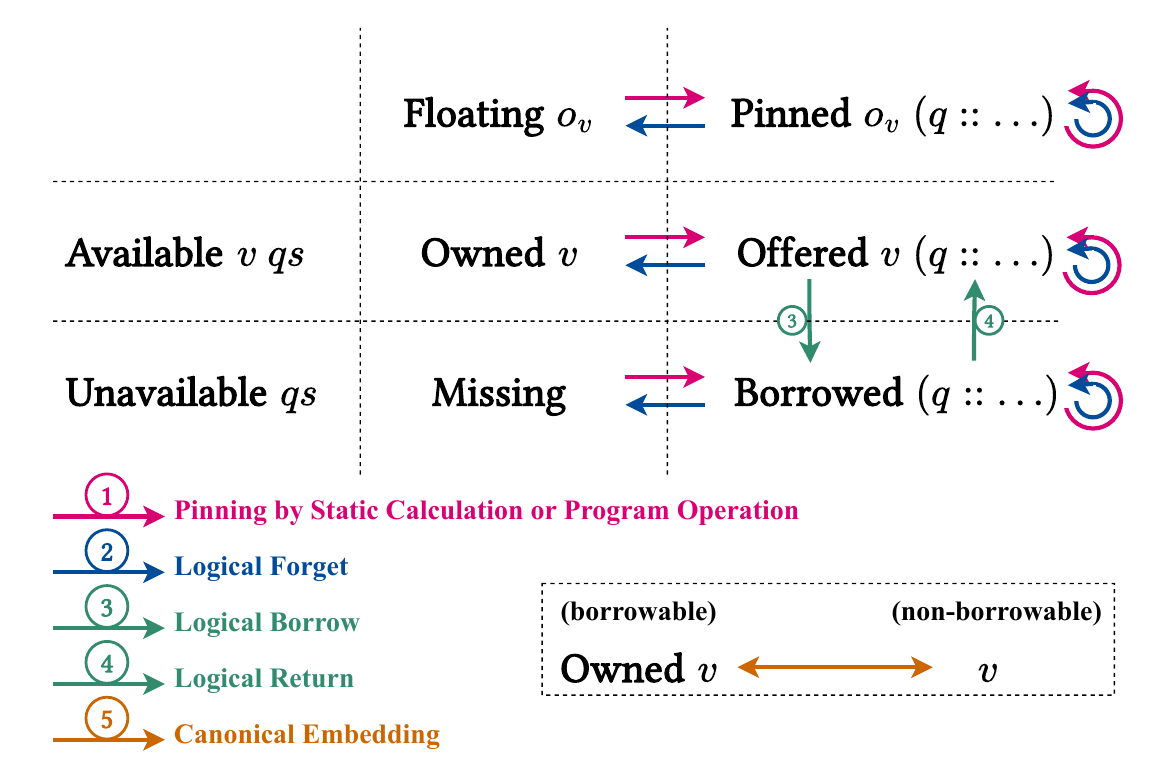}
  }
  \caption{Borrowing states and transitions.}\label{fig:borrowing-states}
  \Description{Borrowing states and transitions.}
\end{figure}

The logical datatype \(\borrowable\) models the borrowing states and lifts a base type \(A\) to its borrow-aware form:
\begin{flalign*}
&\quad \BorrowableFull{A}\; \text{(written \(\Borrowable{A}\))} \eqdef (\Option{A}) \times (\ListType{\loc}) \\
& \begin{array}{l|l|l}
  & \; \Floating{o_v} & \; \Pinned{o_v}{q}{qs} \rule{0pt}{2.5ex}\\
  & \eqdef \tupleTwo{o_v}{\mspace{38mu} \nil} & \eqdef \tupleTwo{o_v}{\mspace{39mu} \Cons{q}{qs}} \\[3pt]
  \hline
  \prelinespace{3pt}
  \; \Available{v}{qs} & \; \Owned{v} & \; \Offered{v}{q}{qs} \\
  \eqdef \tupleTwo{\Some{v}}{qs} & \eqdef \tupleTwo{\Some{v}}{\nil} & \eqdef \tupleTwo{\Some{v}}{\Cons{q}{qs}} \\[3pt]
  \hline
  \prelinespace{3pt}
  \; \Unavailable{qs} & \; \Missing & \; \Borrowed{q}{qs} \\
  \eqdef \tupleTwo{\none}{\mspace{15mu} ps} & \eqdef \tupleTwo{\none}{\mspace{14mu} \nil} & \eqdef \tupleTwo{\none}{\mspace{14mu} \Cons{q}{qs}}
\end{array}
\end{flalign*}

The \(\offered{}\) and \(\borrowed{}\) states capture the fact that the borrowable subpart remains at its previously exposed location \(q\) (with additional aliases \(qs\)); we call them the \textit{pinned states}. Pinned states indicate the validity of exposed internal pointers to the element.

\(\owned{}\) and \(\missing\) are \textit{floating states}. For flat containers such as arrays, vectors and records, floating states can be upgraded to pinned states purely by static calculation (transition \circled{1} in \cref{fig:borrowing-states}), i.e., by computing each subpart’s location from the container’s base address and the subpart's offset that is known statically.

\(\missing{}\) and \(\borrowed{}\) are \textit{unavailable states}, indicating that the subpart has been logically detached from the container and replaced by a tagged hole. In contrast, \textit{available states}, \(\owned\) and \(\offered\), indicate that the container owner has ownership of this subpart.

\subsubsection{Containers with Borrowable Subparts}\label{model:borrowable-containers}

\paragraph{Logical datatypes}
Consider the following standard definition of a logical list. We write \(\nil\) for \(\pred{Nil}\), \(\Cons{x}{l}\) for \(\appTwo{\cons}{x}{l}\), and \([x_1; \dots; x_n]\) for \(\appTwo{\cons}{x_1}{(\dots (\appTwo{\cons}{x_n}{\pred{Nil}}))}\).
\[
\ListType{A} \eqdef \pred{Nil} \;|\; \appTwo{\pred{Cons}}{(x:A)}{(l:\ListType{A})}
\]
By lifting the element type from \(A\) to \(\Borrowable{A}\), we obtain lists whose elements can be borrowed independently:
\[
\ListEType{A} \eqdef \ListType{\Borrowable{A}}
\]
This definition captures the borrowing pattern in which each element of a container can be accessed independently. For instance, in the logger example (\cref{model:step-by-step-guide}), the borrow-aware dictionary has type \(\DictType{\Borrowable{\loggerType}}\).

Supporting borrowable recursive substructures (such as list tails or subtrees) requires modifying the inductive definition of the container (its underlying ``shape''). For example, \(\listSType\) describes lists in which each tail is borrowable:
\[
\ListSType{A} \eqdef \nilSFull \;|\; \ConsSFull{(x:A)}{(\bl:\Borrowable{(\ListSType{A})})}
\]
We write \(\nilS\) for \(\nilSFull\) and \(\ConsS{x}{\bl}\) for \(\ConsSFull{x}{\bl}\).

Combining these two forms, we obtain \(\ListESType{A}\) for lists in which both elements and tails are borrowable:
\[
\ListESType{A} \eqdef \ListSType{\Borrowable{A}}
\]

\paragraph{Representation predicates}

The generic \textit{borrow transformer} on representation predicates, \(\liftToB\), lifts a predicate \(R\) of type \(\ReprType{A}\) to a predicate of type \(\ReprType{\Borrowable{A}}\). We write \(\yB{R}\) for \(\LiftToB{R}\). The transformer is defined as follows:
\begin{flalign*}
  & \mspace{70mu} \bR{\tupleTwo{o_v}{qs}}{p} \eqdef \bvR{o_v}{p} \starsp
\blocR{qs}{p} \\
  & \begin{aligned}
    & \bvR{(\Some{v})}{p} \eqdef \R{v}{p} \qquad \blocR{qs}{p} \eqdef \hpure{\Eqlocs{q}{qs}{p}} \\
    & \bvR{(\none\phantom{\;v})}{p} \eqdef \hpure{\top} \\
  \end{aligned}
\end{flalign*}
The representation predicate \(\yB{R}\) for the borrowable value \(\tupleTwo{o_v}{qs}\) has two components: the value part \(\bvR{o_v}{p}\), which describes the underlying value if available; and the location part \(\blocR{qs}{p}\), which asserts that every pointer in the exposed alias list \(qs\) is equal to the canonical location \(p\).

Tracking an alias list rather than a single optional exposed pointer simplifies specifications: equality among all aliases is encapsulated in \(\yB{R}\), so functions that read or create aliases do not need to track it explicitly. For example, consider \texttt{get} on a dictionary where the queried logger \(g\) already has an exposed pointer \(q\). One could explicitly track the new alias \(r\):
\begin{flalign*}
  & \tstate{ \hpure{d[k] = \OfferedOne{g}{q}} \star \ReprDictE{d}{p}} \\
  &&& \mathllap{
        \codeAppTwo{\texttt{get}}{p}{k}\;
        \tstatex{r}{\hpure{r = q} \star \ReprDictE{d}{p}}
      }
\end{flalign*}
Using the alias list, the same postcondition can be expressed equivalently and more cleanly as (in the same manner as the specifications in \cref{model:step-by-step-guide}):
\begin{flalign*}
  & \tstatex{r}{\ReprDictE{d[k := \OfferedTwo{g}{r}{q}]}{p}} \\
  \equiv\; & \tstatex{r}{\ReprDictE{d[k := \Pin{d[k]}{r}]}{p}}
\end{flalign*}

The borrow transformer ``\(\yB{}\)'' enables modular lifting of representation predicates to support borrow tracking. Starting from the standard list representation predicate \(\reprList\) (\cref{sec:introduction}), we define \(\reprListE\) that supports borrowing elements:
\[
\reprListE: (\ReprType{A}) \rightarrow \ReprType{(\ListEType{A})} \quad
\appOne{\reprListE}{R} \eqdef \appOne{\reprList}{\Borrowable{R}}
\]
A second variant, \(\reprListS\), supports borrowing tails:
\begin{flalign*}
  & \reprListS: (\ReprType{A}) \rightarrow \ReprType{(\ListSType{A})} \\
  & \ReprListS{R}{\nilS}{p} \eqdef \hpure{p = \nullptr} \\
  & \ReprListS{R}{(\ConsS{x}{\bl})}{p} \eqdef \R{x}{p} \star\, \hexistsx{q} \pointsto{p+1}{q} \star \bReprListS{R}{\bl}{q} \end{flalign*}
Similarly, \(\reprListES\) combines both forms, supporting borrowable elements and borrowable tails:
\[
\reprListES: (\ReprType{A}) \rightarrow \ReprType{(\ListESType{A})} \quad
\appOne{\reprListES}{R} \eqdef \appOne{\reprListS}{\Borrowable{R}}
\]

\subsubsection{Borrowing State Transitions}\label{model:borrowing-state-transitions}

\Cref{fig:borrowing-states} shows the transitions between the borrowing states.

\textit{\textbf{Transition \circled{1} (Pinning by Static Calculation or Program Operation)}} pins a container subpart at a newly exposed location. Static calculation, as already discussed in \cref{model:borrowing-states}, derives the subpart's location based on static knowledge. It is applicable only in flat containers.

For most pointer-based containers, subpart locations are often not statically known, and must instead be computed dynamically by executing traversal operations. For example, in the logger example, the \texttt{get} function exposes the location of a logger stored in a linked list, whose address cannot be determined statically.

A pinning transition always produces a pinned state. Assume that the newly exposed location is \(r\). Such a transition can convert a floating state into a pinned state, for example, from \(\Owned{v}\) to \(\OfferedOne{v}{r}\). If starting from an already pinned state such as \(\Offered{v}{q}{qs}\), it extends the alias list with \(r\), yielding \(\Offered{v}{r}{\Cons{q}{qs}}\).

\textit{\textbf{Transition \circled{2} (Logical Forget)}} is the reverse of pinning: it logically invalidates an exposed pointer by discarding the pure constraint that tracks it.
\[
  \textit{\small (logical forget)}\quad
  \bR{\tupleTwo{o_v}{\Cons{q}{qs}}}{p} \vdash \bR{\tupleTwo{o_v}{qs}}{p}
\]

Logical forget always starts with a pinned state. If all pointers in the alias list are forgotten, it yields a floating state; otherwise the produced state is still pinned. Logical return is useful for unifying a state with one without certain aliases, such as postconditions that do not track pointers exposed temporarily within the function body.

\textit{\textbf{Transition \circled{3} (Logical Borrow)}} and
\textit{\textbf{Transition \circled{4} (Logical Return)}} implement the key feature of our model: supporting flexible ownership transfers.

Specifically, a subpart offered at an exposed location can be logically borrowed from the container for local reasoning. Its ownership can be later reclaimed by the original container, as long as the subpart is not moved (but it may be mutated). Formally, the logical borrow and return rules are captured by the following equivalence:
\[
\appTwoTight{\yB{R}}{(\Offered{v}{q}{qs})}{p} \equiv \appTwoTight{\yB{R}}{(\Borrowed{q}{qs})}{p} \star \appTwo{R}{v}{q}
\]
\Cref{fig:offered-borrowed} provides an intuitive illustration. A container \(c\) with elements \(e_1\) offered at \(p_1\) and \(e_3\) offered at \(p_3\) can be decomposed into (1) a partial container with holes at \(p_1\) and \(p_3\), and (2) the two detached elements \(e_1\) and \(e_3\). This decomposition corresponds to logical borrow, and the reverse operation (recomposition) to logical return.

\textit{\textbf{Transition \circled{5} (Canonical Embedding)}} shows that
non-borrowable values are equivalent to owned borrowable value:
\[
  \R{v}{p} \equiv \bR{(\Owned{v})}{p}
\]
We often omit this transition in the rest of the paper.

\subsection{Indirection}\label{model:indirection}

Many containers store their elements or substructures \textit{indirectly} through intermediate pointers rather than embedding them in place. Such indirection is standard in pointer-based containers such as linked lists and trees.

Unlike borrowability, modeling indirection does not require modifying the logical datatype, as functional datatypes already abstract away low-level pointer indirection.

We introduce the \textit{indirection transformer} ``\(\yI{}\)'' to map a representation predicate \(R\) for values stored in place to a predicate for values stored via an intermediate pointer. Although ``\(\yI{}\)'' is not directly related to borrowing, it composes nicely with the borrow transformer ``\(\yB{}\)''.
\begin{flalign*}
  \yI{}: \forallx{A} (\ReprType{A}) \rightarrow \ReprType{A}
  \quad \iR{v}{p} \eqdef \hexistsx{q} \pointsto{p}{q} \starsp \R{v}{q}
\end{flalign*}
With ``\(\yI{}\)'', the non-empty case of \(\reprList\) can be written as:
\[
\ReprList{R}{(\Cons{x}{l})}{p} \equiv \R{x}{p} \starsp \iReprList{R}{l}{(p+1)}
\]

\subsection{Compositions of Predicate Transformers}\label{model:compositions}

The two representation predicate transformers ``\(\yB{}\)'' and ``\(\yI{}\)'' serve as the fundamental abstraction for modeling recursive containers with borrowable subparts. By composing them, we can express different patterns of memory organization. Here, ``\(\circ\)'' denotes the standard function composition operator: \(\appOne{(g \circ f)}{x} = \appOne{g}{(\appOne{f}{x})}\).

\(\yI{} \circ \yB{}\) models a borrowable value stored indirectly. For example, the non-empty case of \(\reprListS\) can be written as:
\[
\ReprListS{(\ConsS{x}{\bl{}})}{p} \equiv \R{x}{p} \starsp \ibReprListS{R}{\bl}{(p+1)}
\]
Here, \(\ibReprListS{R}{\bl}{(p+1)}\) states that an intermediate pointer is stored at \(p+1\), and this pointer points to the borrowable tail \(\bl{}\). This predicate is useful for reasoning about in-place updates to the tail, but not about modifying the intermediate pointer, since it is not borrowable.

\(\yB{} \circ \yI{}\) models the case where the intermediate pointer itself can be borrowed (along with the memory it points to). Consider the lists modeled by \(\reprListSPrime\) which is defined as follows:
\begin{flalign*}
& \reprListSPrime: (\ReprType{A}) \rightarrow \ReprType{(\ListSType{A})} \\
& \ReprListSPrime{R}{\nilS}{p} \eqdef \hpure{p = \nullptr} \\
& \ReprListSPrime{R}{(\ConsS{x}{\bl})}{p} \eqdef{}
\R{x}{p} \starsp \biReprListSPrime{R}{\bl}{(p+1)}
\end{flalign*}
Here, \(\biReprListSPrime{R}{\bl}{(p+1)}\) states that a borrowable intermediate pointer is stored at \(p+1\), and borrowing this pointer transitively borrows the tail it points to. This predicate is useful when the client may update the tail by modifying the intermediate pointer, for example, to point to a different list.

We have \(\yIB{R} ~\neq~ \yBI{R}\), \(\yII{R} ~\neq~ \yI{R}\), but
\begin{flalign*}
  \bbR{\tupleTwo{\Some{\bv{}}}{qs}}{p} &~=~ \bR{\bv{}}{p} \starsp \bblocR{qs}{p} \\
  \bbR{\tupleTwo{\none}{qs}}{p} &~=~ \bblocR{qs}{p}
\end{flalign*}

By composition of ``\(\yB{}\)'' and ``\(\yI{}\)'', we can express a wide range of memory manipulation patterns in a principled and reusable way. For example, \(\yBIB{R}\) is useful when the intermediate pointer and pointed-to object are both borrowable.

\section{Case Studies}\label{sec:case-studies}

\note{We avoid orthogonal challenges of C such as fixed-size integers, byte representations and alignment.}

In this section, we illustrate logical pinning across a series of containers. We adopt a simple imperative \(\lambda\)-calculus with dynamic allocation, along with the linear separation logic presented in \cref{bg:separation-logic}. The language includes let-bindings, conditional expressions, fixpoints, equality checks, pointer arithmetic, and mutable arrays. Its core heap-manipulating primitives are as follows:
\begin{flalign*}
\quad
& \Alloc{n} && \text{\begin{tabular}[t]{@{}l@{}}
                      Allocate \(n\) contiguous uninitialized cells, \\
                      and return the starting location.
               \end{tabular}} \\
& \Free{n}{p} && \text{Deallocate \(n\) contiguous cells starting at \(p\).} \\
& \assign{p}{v} && \text{Write value \(v\) to memory location \(p\).} \\
& !p && \text{Read value at memory location \(p\).}
\end{flalign*}

Our development, including correctness proofs of container implementations with respect to our specifications and case-study programs, is formalized on top of a variant of the core CFML calculus in the Rocq proof assistant. The formalization is available at:
\begin{center}
\url{https://github.com/epfl-systemf/logical-pinning}
\end{center}


\subsection{Flat Containers}\label{case-studies:flat-containers}

\subsubsection{Single Cells}\label{case-studies:single-cells}

We present two examples of cells that, while not useful in isolation as cells are the smallest memory units, become valuable if combined into larger structures.


\paragraph{A cell with a borrowable primitive value} Such a cell can be read from or written to when the borrowable value is available. Neither operations invalidate pointers to this cell:
\begin{flalign*}
& \tstate{\bCell{(\Available{v}{qs})}{p}}\; !p\; \\
&&& \mathllap{
    \tstatex{r}{\hpure{r=v} \star \bCell{(\Available{v}{qs})}{p}}
    } \\
& \tstate{\bCell{(\Available{v}{qs})}{p}}\;
  \assign{p}{u}\; \\
&&& \mathllap{
    \tstate{\bCell{(\Available{u}{qs})}{p}}
  }
\end{flalign*}

\paragraph{A cell with an indirectly-stored borrowable value} Assume that the cell at location \(p\) is described by \(\iR{v}{p}\), that is, it stores an intermediate pointer to \(v\), and \(v\) can be described by predicate \(R\). Reads and writes to this cell operate on the intermediate pointer without affecting \(v\):
\begin{flalign*}
& \triplex{!p}{\iR{v}{p}}{r}{\Cell{p}{r} \starsp \R{v}{r}} \\
& \tripleUnit{\assign{p}{q}}{\iR{v}{p}}{\Cell{p}{q} \starsp  \hexistsx{q_v} \R{v}{q_v}}
\end{flalign*}
Although the above specification is precise, it exposes the internal structure of ``\(\yI{}\)'' and thus breaks the abstraction of indirection, reducing modularity. For example, it does not allow reading from or writing to the same cell twice, since the pre- and postconditions do not have the same shape.

The intermediate pointer being operated on is an internal pointer of this single-cell container, thereby its exposure or mutation can instead be handled by our borrowing model:
\begin{flalign*}
& \triplex{!p}{\ibR{\bv{}}{p}}{r}{\ibR{(\Pin{\bv{}}{r})}{p}} \\
\equiv\; &\triplex{!p}{\ibR{\bv{}}{p}}{r}{\ibR{(\BorrowedOne{r})}{p} \star \bR{\bv{}}{r}} \\
& \tstate{\ibR{(\BorrowedOne{r})}{p} \star \bR{\bv{}}{r}}\;
  \assign{p}{q}\; \\
&&& \mathllap{
    \tstateUnit{\ibR{(\BorrowedOne{q})}{p} \starsp \bR{\bv}{r}}
    }
\end{flalign*}
Here, we make the inner value borrowable. Exposing the intermediate pointer pins the inner value, and replacing this pointer with \(q\) is treated as replacing the inner value with a hole at \(q\) (the target of \(q\) is unknown).

\subsubsection{Arrays and Records}\label{case-studies:arrays-and-records}

A plain array is a contiguous block of memory without a header as in C, and we model it as a list of values laid out sequentially in memory:
\begin{flalign*}
& \arrayType:\forallx{A} (\ReprType{A}) \rightarrow \ReprType{(\ListType{A}}) \\
& \ReprArray{R}{\nil{}}{p} \eqdef \hpure{p = \nullptr} \\
& \ReprArray{R}{(\Cons{x}{l})}{p} \eqdef \hpure{p \neq \nullptr} \star \R{x}{p} \star \ReprArray{R}{l}{(p+1)}
\end{flalign*}

We consider records as \textit{heterogeneous} arrays, where each field occupies one cell that stores a value in place or an intermediate pointer to the value, and the representation predicate of each field may differ. To uniformly describe such fields, we introduce a helper type \(\reprval\) to package a value with its representation predicate:
\[
  \reprval \eqdef \mkReprValFull{(t: \type)}{(R_t:\ReprType{t})}{(v: t)}
\]
When \(t\) is clear from context, we write \(\mkReprVal{R}{v}\) with the type left implicit.

\todo{rename ``offset''?}

Each field label \(f\) is associated with a natural number index \(\offset{f}\), indicating the field's offset from the entry location of the record. The representation predicate for records is defined as follows:
\begin{flalign*}
  & \record: \ReprType{(\ListType{(\fieldType \times \reprval)})} \\
  & \Record{\nil{}}{p} \eqdef \hpure{p = \nullptr} \\
  & \Record{(\Cons{\tupleTwo{f}{\mkReprVal{R}{v}}}{l})}{p} \eqdef \\
  & \quad \hpure{p \neq \nullptr} \star \appTwo{R}{v}{(p+\offset{f}) \star \Record{l}{p}}
\end{flalign*}
We write
\(
  \mapThree
    {\field{\(f_0\)}{R_0}{v_0}}
    {\dots}
    {\field{\(f_{n-1}\)}{R_{n-1}}{v_{n-1}}}
  \)
for a record with \(n\) fields \(f_0\) to \(f_{n-1}\). For each \(i \in \{0, \dots, n-1\}\), field \(f_i\) stores value \(v_i\) with representation predicate \(R_i\).

Records serve as a fundamental abstraction for defining containers in our approach. Recall the list representation predicate \(\reprListS\) (\cref{model:borrowable-containers}), its non-empty case can be encoded as a record with two fields:
\renewcommand{\content}{\recordListTight{R}{x}{\ibReprR{\reprList}{R}}{l}}
\[ \ReprListS{R}{(\ConsS{x}{l})}{p} \equiv \Record{\content}{p} \]
where \(\offset{\fieldname{elem}}\) is \(0\), \(\offset{\fieldname{next}}\) is \(1\).

We add the following notations as syntactic sugar for field access and exposing field location:

\noindent\centerline{\begin{tabular}{lll}
\toprule
\textbf{Notation} & \textbf{C Syntax} & \textbf{Desugars to} \\
\midrule
\(\readField{p}{f}\) & \texttt{(*p).f}, \texttt{p{\textrightarrow}f} & \(! (p + \offset{f})\) \\
\(\writeField{p}{f}{v}\) & \texttt{(*p).f = v}, \texttt{p{\textrightarrow}f = v} & \(\assign{(p + \offset{f})}{v}\) \\
\(\locField{p}{f}\) & \texttt{\&(p\textrightarrow f)} & \(p + \offset{f}\) \\
\bottomrule
\end{tabular}}

\paragraph{Field access}
Accessing a record field reduces to accessing a single cell (\cref{case-studies:single-cells}), hence we can lift the single-cell specifications for field operations. For example, reading a field that stores a borrowable value in place is specified as follows. Crucially, only the target field’s value must be available, and other fields may be borrowed independently.
\begin{flalign*}
  & \tstate{\hpure{l[f] = \mkReprVal{\bcell}{(\Available{v}{qs})}} \star \Record{l}{p}}\; \\
  &&& \mathllap{
      \readField{p}{f}\;
      \tstatex{r}{\hpure{r=v} \star \Record{l}{p}}
      }
\end{flalign*}

\paragraph{Exposing field location} Because records are flat containers, a field can always be pinned at its address statically:
\begin{flalign*}
  & \Record{\subst{l}{f}{\mkReprVal{\yB{R}}{\bv{}}}}{p} \\
  \equiv\; & \Record{\subst{l}{f}{\mkReprVal{\yB{R}}{(\Pin{\bv{}}{(p+\offset{f})})}}}{p}
\end{flalign*}
Alternatively, the location can be obtained dynamically:
\begin{flalign*}
  & \tstate{\hpure{l[f] = \mkReprVal{\yB{R}}{\bv{}}} \star \Record{l}{p}}\; \locField{p}{f}\; \\
  &&& \mathllap{
      \tstatex{r}
        {\Record{\subst{l}{f}{\mkReprVal{\yB{R}}{(\Pin{\bv}{r})}}}{p}}
      }
\end{flalign*}
By the definition of \(\record\), \(r\) equals \(p+\offset{f}\).

\subsection{Recursive Containers}\label{case-studies:recursive-containers}

Recursive containers are the natural use case of the logical-pinning model: element locations are not statically known, and their pointer-rich APIs often expose internal pointers.  Our model marks these pointers on demand; we demonstrate its generality and practicality on linked lists and trees, by deriving precise specifications for standard APIs.

{
\setlength{\abovecaptionskip}{5pt}
\begin{figure*}[htbp]
\[\arraycolsep=2pt\def\arraystretch{1}
\begin{array}{rcl}
\tstate{\ReprCListE{l}{p}}
&\codeAppOne{\texttt{size}}{p}
&\tstatex{r}{\hpure{r = \Length{l}} \star \ReprCListE{l}{p}}
\\
\tstate{\ReprCListE{l}{p}}
&\codeAppOne{\texttt{is\_empty}}{p}
&\tstatex{r}{\hpure{r=\valTrue{} \leftrightarrow l=\nil} \star \ReprCListE{l}{p}}
\\
\tstate{\ReprCListE{(\Cons{(\Available{v}{qs})}{l})}{p}}
&\readField{p}{\fieldname{elem}}
&\tstatex{r}{\hpure{r=v} \star \ReprCListE{(\Cons{(\Available{v}{qs})}{l})}{p}}
\\
\tstate{\ReprCListE{(\Cons{(\Available{v}{qs})}{l})}{p}}
&\writeField{p}{\fieldname{elem}}{u}
&\tstateUnit{\ReprCListE{(\Cons{(\Available{u}{qs})}{l})}{p}}
\\
\tstate{\hpure{\forallx{\bv \in l} \isAvailable{\bv}} \star \ReprCListE{l}{p}}
&\codeAppOne{\texttt{incr\_all}}{p}
&\tstatex{r}{\ReprCListE{(\MapE{(\lambdax{v} v+1)}{l})}{p}}
\\
\tstate{\hpure{n < \Length{l} \wedge \Nth{n}{l}=\Available{v}{qs}} \star \ReprCListE{l}{p}}
&\codeAppTwo{\texttt{nth\_elem}}{n}{p}
&\tstatex{r}{\hpure{r=v} \star \ReprCListE{l}{p}}
\\
\tstate{\hpure{n < \Length{l}} \star \ReprCListE{l}{p}}
&\codeAppTwo{\texttt{nth\_elem\_ptr}}{n}{p}
&\tstatex{r}{\ReprCListE{\subst{l}{n}{\Pin{(\Nth{n}{l})}{r}}}{p}}
\\
\tstate{\ReprCListE{(\Cons{\bv}{l})}{p}}
&\codeAppOne{\texttt{pop\_front}}{p}
&\tstatex{r}{\ReprCListE{l}{r} \star \bCell{\bv}{p}}
\end{array}
\]
\caption{API specification for lists with borrowable elements and non-borrowable tails.}\label{fig:spec-list-borrowable-elements}
\Description{API specification for lists with borrowable elements and non-borrowable tails.}
\end{figure*}

\begin{figure*}
\[\arraycolsep=2pt\def\arraystretch{1}
\begin{array}{rcl}
\tstate{\hpure{\sForall{(\Owned{l})}} \star \ReprCListES{l}{p}}
&\codeAppOne{\texttt{size}}{p}
&\tstatex{r}{\hpure{r = \bbLength{l}} \star \ReprCListES{l}{p}}
\\
\tstate{\ReprCListES{l}{p}}
&\codeAppOne{\texttt{is\_empty}}{p}
&\tstatex{r}{\hpure{r=\valTrue{} \leftrightarrow l=\nilS} \star \ReprCListES{l}{p}}
\\
\tstate{\ReprCListES{(\ConsS{(\Available{v}{qs})}{\bl})}{p}}
&\readField{p}{\fieldname{elem}}
&\tstatex{r}{\hpure{r=v} \star \ReprCListES{(\ConsS{(\Available{v}{qs})}{\bl})}{p}}
\\
\tstate{\ReprCListES{(\ConsS{(\Available{v}{qs})}{\bl})}{p}}
&\writeField{p}{\fieldname{elem}}{u}
&\tstateUnit{\ReprCListES{(\ConsS{(\Available{u}{qs})}{\bl})}{p}}
\\
\tstate{\hpure{\esForall{(\Owned{l})}} \star \ReprCListES{l}{p}}
&\codeAppOne{\texttt{incr\_all}}{p}
&\tstatex{r}{\ReprCListES{(\MapES{(\lambdax{v} v+1)}{l})}{p}}
\\
\makecell[vr]{
  \begin{mstate}
    \hpure{n < \bbLength{l} \wedge \NthES{n}{l} = \Available{v}{qs}} & \\
    \starsp \ReprCListES{l}{p} &
  \end{mstate}
}
&\codeAppTwo{\texttt{nth\_elem}}{n}{p}\;
&\tstatex{r}{\hpure{r=v} \star \ReprCListES{l}{p}}
\\
\tstate{\hpure{n < \bbLength{l}} \star \ReprCListES{l}{p}}
&\codeAppTwo{\texttt{nth\_elem\_ptr}}{n}{p}
&\tstatex{r}{\ReprCListES{\subst{l}{n}{\Pin{(\NthES{n}{l})}{r}}}{p}}
\\
\tstate{\ReprCListES{(\ConsS{\bv}{\Available{l}{qs}})}{p}}
&\codeAppOne{\texttt{pop\_front}}{p}
&\tstatex{r}{\ReprCListES{l}{r} \star \hpure{\Eqlocs{q}{qs}{r}} \star \bCell{\bv}{p}}
\\
\tstate{\ReprCListES{(\ConsS{\bv{}}{\bl})}{p}}
&\new{\readFieldNext{p}}
&\tstatex{r}{\ReprCListES{(\ConsS{\bv{}}{\Pin{\bl}{r}})}{p}}
\\
\tstate{\ReprCListES{(\ConsS{\bv{}}{\bl})}{p}}
&\new{\writeFieldNext{p}{q}}
&\tstatex{r}{\ReprCListES{(\ConsS{\bv{}}{\BorrowedOne{q}})}{p} \star \hexistsx{q_{\bl{}}} \bReprCListES{\bl{}}{q_{\bl{}}}}
\\
\tstate{\hpure{n < \bbLength{l}} \star \ReprCListES{l}{p}}
&\new{\codeAppTwo{\texttt{nth\_tail}}{n}{p}}
&\tstatex{r}{\ReprCListES{(\SubstTailES{n}{(\Pin{(\SkipnES{n}{l})}{r})}{l})}{p}}
\\
\tstate{\ReprCListES{l}{p}}
&\new{\codeAppTwo{\texttt{push\_front}}{v}{p}}
&\tstatex{r}{\ReprCListES{(\ConsS{(\Owned{v})}{\OfferedOne{l}{p}})}{r}}
\\
\makecell[vr]{
  \begin{mstate}
    \hpure{\sForall{(\Owned{l_1})}} & \\
    \starsp \ReprCListES{l_1}{p_1} \star \ReprCListES{l_2}{p_2} &
  \end{mstate}
  }
&\new{\codeAppTwo{\texttt{append}}{p_1}{p_2}}
&\tstatex{r}{\bReprCListES{(\bbAppend{(\OfferedOne{l_1}{p_1})}{(\OfferedOne{l_2}{p_2})})}{r}}
\\
\tstate{\hpure{\sForall{(\Owned{l})}} \star \ReprCListES{l}{p}}
&\new{\codeAppTwo{\texttt{push\_back}}{v}{p}}
&\tstatex{r}{\bReprCListES{(\bbAppend{(\OfferedOne{l}{p})}{\ListOwnedTail{\Owned{v}}})}{r}}
\end{array}
\]
\caption{API specification for lists with borrowable elements and borrowable tails. \(\sForall{\bl{}}\) (resp. \(\esForall{\bl{}}\)) asserts every tail (resp. subpart) of \(\bl{}\) is available; \(\bbAppend{\blOne{}}{\blTwo{}}\) denotes the borrow-aware concatenation of two lists; \(\ListOwnedTail{x}\) denotes \(\ConsS{x}{\Owned{\nilS}}\).}\label{fig:spec-list-with-borrowable-tail}
\Description{API specification for lists with borrowable elements and borrowable tails.}
\end{figure*}
}

\subsubsection{Linked Lists}\label{case-studies:linked-lists}

We study singly linked lists, with nodes represented in C as \texttt{lnode} (\cref{sec:introduction,model:borrowable-containers}). We start with the traditional representation of lists --- a logical model whose elements and tails are not borrowable:
\[
\reprCList: \ReprType{(\ListType{\intType})} \qquad
\reprCList \eqdef \appOne{\reprList}{\cell}
\]
Consider the specification of the function \texttt{size}, which computes the list's length \(\Length{l}\), in this model:
\[
\triplex{\codeAppOne{\texttt{size}}{p}}{\ReprCList{l}{p}}{r}{\hpure{r = \Length{l}} \star \ReprCList{l}{p}}
\]
Even for such a simple function, the \(\reprCList\)-base specification is imprecise. \texttt{size} neither exposes internal pointers, nor relocates objects, nor transfers ownership, yet (1) it demands more permissions than necessary, as \texttt{size} depends only on the list shape and does not need ownership of elements; and (2) it does not capture pointer stability, which makes pointer tracking impossible.

Let us instead use logical pinning.  We introduce \(\reprCListE\), which describes the same list but with borrowable elements.  This lets specifications require only the necessary element ownership, and track pointers to elements:
\[
\reprCListE: \ReprType{(\ListType{\Borrowable{\intType}})} \qquad
\reprCListE \eqdef \appOne{\reprList}{\bcell}
\]
\Cref{fig:spec-list-borrowable-elements} shows a complete \(\reprCListE\)-based API specification.  In particular, the new specification of \texttt{size} permits elements to be unavailable, preserves their borrowing states, and therefore preserves all exposed pointers tracked in those states.  Other functions are \texttt{is\_empty}, which returns \(\valTrue\) iff the list is empty;  \(\readFieldElem{p}\) and  \(\writeFieldElem{p}{u}\), which read and write the ``\(\fieldname{elem}\)'' field; \texttt{incr\_all}, which increments all elements in place; \texttt{nth\_elem} and \texttt{nth\_elem\_ptr}, which return the value of, and a pointer to, the \(n\)-th element; and \texttt{pop\_front}, which removes the head and returns a pointer to the new head.

These \(\reprCListE\)-based specifications are more precise than \(\reprCList\)-based ones: they require minimal element ownership, and account for all possible exposed pointers to elements.  For example, the specification of \texttt{nth\_elem\_ptr} tracks the returned element pointer by simply pinning it.

Some APIs also expose pointers to tails, rather than just elements. For those, we introduce \(\reprCListES\), which makes both elements and tails borrowable:
\[
\reprCListES: \ReprType{(\ListSType{\Borrowable{\intType}})} \qquad
\reprCListES \eqdef \appOne{\reprListS}{\bcell}
\]
\Cref{fig:spec-list-with-borrowable-tail} shows the \(\reprCListES\)-based API, with \new{operations that expose tail pointers} highlighted. Such operations are difficult to specify precisely with \(\reprCListE\), and therefore do not appear in \cref{fig:spec-list-borrowable-elements}. They include operations on the \texttt{next} field, \texttt{push\_front} (which adds a new head), \texttt{append} (which concatenates a second list), and \texttt{push\_back} (which appends a new node at the tail).

These \(\reprCListES\)-based specifications are \textit{maximally} precise: they account for every internal pointer to elements and tails, and require minimal ownership.  For example, the specification of \texttt{append} preserves the two original head pointers \(p_1\) and \(p_2\) in a clean and natural form.  All \(\reprCListE\)-based specifications can be derived from the \(\reprCListES\)-based ones.

\paragraph{Case study: swaps}\label{case:element-swap}
Our specification of \texttt{nth\_elem\_ptr} requires no element ownership, and tracks exposed pointers without altering the container representation.  This lets us prove the correctness of an element-swapping program that invokes \texttt{nth\_elem\_ptr} twice, on the same or distinct indices, to obtain element pointers, and then calls \texttt{memswap} to exchange their contents (\cref{app:full-swap-proof}). Here too, we see that logical pinning naturally supports traditionally hard-to-verify patterns.

\subsubsection{Binary Trees}\label{case-studies:binary-trees}

We define \(\treeSType\) with predicate \(\reprTreeS\) to model binary trees with borrowable subtrees:
\renewcommand{\content}{
  \recordTree{R}{x}
    {\ibReprR{\reprTreeS}{R}}{\btreel}
    {\ibReprR{\reprTreeS}{R}}{\btreer}{}
}
\begin{flalign*}
  &\TreeSType{A} \eqdef \leafS \;|\; \appTwo{\nodeS}{(x:A)}{(\btreel, \btreer: \Borrowable{(\TreeSType{A}))}} \\
  & \reprTreeS: (\ReprType{A}) \rightarrow \ReprType{(\TreeSType{A})} \\
  & \ReprTreeS{R}{\leafS}{p} \eqdef \hpure{p = \nullptr} \\
  & \ReprTreeS{R}{(\NodeS{x}{\btreel}{\btreer})}{p} \eqdef \Record{\\&\quad \content}{p}
\end{flalign*}
In contrast, a custom representation predicate \(\reprHoleCutTree\) for trees with holes and cuts (\cref{bg:trees-with-holes-and-cuts}) would be written as:
\renewcommand{\content}[2]{
  \recordTree{#1}{x}
    {\iReprR{\reprHoleCutTree}{R}}{\btreel}
    {\iReprR{\reprHoleCutTree}{R}}{\btreer}{#2}
}
\begin{flalign*}
  & \reprHoleCutTree: (\ReprType{A}) \rightarrow \ReprType{(\HoleCutTree{A})} \\
  & \ReprHoleCutTree{R}{\leaf}{p} \eqdef \hpure{p = \nullptr} \\
  & \ReprHoleCutTree{R}{(\Node{x}{t_l}{t_r})}{p} \eqdef \Record{\\&\; \content{R}{\mspace{30mu}}}{p} \\
  & \ReprHoleCutTree{R}{(\Hole{q}{t_l}{t_r})}{p} \eqdef \Record{\\&\; \content{(=p)}{}}{p} \\
  & \ReprHoleCutTree{R}{(\Cut{q})}{p} \eqdef \hpure{p = q}
\end{flalign*}
Our model subsumes this technique: our type \(\TreeSType{\Borrowable{A}}\) describes trees with \emph{both} borrowable elements and subtrees, unifying holes and cuts as just \(\borrowed\). This lets us reproduce the motivating example of the original paper on trees with holes and cuts~\cite{charguéraudHigherorderRepresentationPredicates2016}, a \texttt{find} function that returns a subtree at a given path, with an enhanced specification that supports repeated lookups and guarantees pointer stability:
\begin{flalign*}
  & \tstate{\bReprTreeS{R}{\bt{}}{p} \star \hpure{\Path{i}{\bt{}}{\btOne{}}}}\;
    \codeAppTwo{\texttt{find}}{i}{p}\; \\
  &&& \mathllap{
        \tstatex{r}{\bReprTreeS{R}{\substb{\bt{}}{i}{\Pin{\btOne{}}{r}}}{p}}
      }
\end{flalign*}
Here, \(\Path{i}{\bt{}}{\btOne{}}\) states that a valid path \(i\) in tree \(\bt{}\) reaches a subtree \(\btOne{}\), and \(\substb{\bt{}}{i}{\bvar{t_x}}\) denotes the borrow-aware substitution of \(\bt{}\) in which the subtree at \(i\) is replaced by \(\bvar{t_x}\).

\paragraph{Case study: binary-tree left rotation}\label{case:tree-rotation}

In this case study, we show that our model simplifies some complex proofs and enables more precise specifications.

During a binary-tree left rotation (\cref{fig:tree-left-rotation}), a subtree is temporarily shared between two trees. In the traditional approach, this situation requires unfolding the tree abstraction two layers deep and managing the resulting split nodes and subtrees, which is conceptually cumbersome. An intermediate state has the following form, where \(\reprTree\) is a traditional representation predicate of binary trees:
\[
  \begin{mstate}
    & \pointsto{p}{x} \starsp \iReprTree{R}{t_l}{(p+1)} \starsp \pointsto{p+2}{p_r} \\
    & \starsp \pointsto{p_r}{x_r} \starsp \pointsto{p_r+1}{p_{rl}} \starsp \iReprTree{R}{t_{rr}}{(p_r+2)} \\
    &  \starsp \hpure{\Eqlocs{q}{qs_r}{p_r}} \starsp \iReprTree{R}{t_{rl}}{p_{rl}}\\
  \end{mstate} \\
\]

With our model, the tree abstraction does not need to be unfolded. The key idea is to represent this intermediate state by letting one tree keep ownership of the subtree and offer it at location \(q\), while the other tree treats this subtree as borrowed at \(q\). The ownership of the shared subtree can then be flexibly transferred between the two trees. Crucially, without the \(\offered\) state, as in trees with holes and cuts, such ownership transfer would not be possible since the tree currently owning the subtree would hide its location.

In the proof using our model (the full proof is given in \cref{app:full-tree-rotation-proof}), this state is represented as:
\[
 \begin{mstate}
  & \ReprTreeS{R}{(\NodeS{x}{\btreel}{(\Borrowed{p_r}{qs_r})})}{p} \\
  &\quad \star \ReprTreeS{R}{(\NodeS{x_r}{(\Pin{\btreerl}{p_{rl}})}{\btreerr})}{p_r}
  \end{mstate} \\
\]

Assume \texttt{left\_rotate} returns a pointer to the root of the rotated tree. Our following specification preserves the original root pointer \(p\). Here, \(\leftRotate\) maps a tree to its rotated form.
\begin{flalign*}
&\tstate{\hpure{t = \NodeS{x}{\btreel}{(\Available{t_r}{qs_r})}} \star \ReprTreeS{R}{t}{p}} \\
& \qquad \texttt{left\_rotate}(p) \\
& \tstatex{r}{\bReprTreeS{R}{(\LeftRotate{(\OfferedOne{t}{p})})}{r}}
\end{flalign*}

\section{Discussion}\label{sec:discussion}

Our logical-pinning model captures ad hoc proof patterns that occur in existing separation-logic developments: (1) the notion of holes and cut in trees (\cref{bg:trees-with-holes-and-cuts}) can be expressed uniformly as \(\pred{Borrowed}\) in our model; (2) when two pointers \(p\) and \(q\) are aliases in separation logic, the common approach is to introduce a pure constraint \(p = q\) and shift ownership between the aliasing pointers via equality rewriting:
\[
\R{v}{p} \star \hpure{p = q} \equiv \R{v}{q} \star \hpure{p = q}
\]
In our model, pinning captures such aliasing constraints:
\[
  \bR{\tupleTwo{o_v}{\Cons{q}{qs}}}{p} \equiv \bR{\tupleTwo{o_v}{\Cons{p}{\Cons{q}{qs}}}}{p} \equiv
  \bR{\tupleTwo{o_v}{\Cons{p}{qs}}}{q}
\]

We have demonstrated that our model enables more precise specifications. We verify the correctness of the logger example using a precise specification of \setlevelall{} that guarantees the function does not relocate elements. For linked lists, we provide API specifications that precisely track all exposed pointers to elements and tails, enabling the verification of representative programs such as element-swapping.

Our model also simplifies proofs that are conceptually complex under traditional approaches. For instance, our proof of tree rotation avoids unfolding the tree representation predicate, keeping the reasoning simple: at most two trees need to be tracked, rather than two split records (each with three fields) and three subtrees.

Our model further enables proving program patterns not supported by traditional specifications, notably \textit{expose-then-mutate}: exposing internal pointers, optionally performing container operations that do not invalidate them, and later using those pointers to read or modify container subparts. This is exactly the pattern in the previous examples (logger, element-swapping, subtree-compare).

\paragraph{Internal sharing within a container}\label{discussion:internal-sharing}

Some containers allow the same subpart to be referenced multiple times within the container. For instance, the same object may be pushed to a queue more than once. In optimized trees such as compressed tries (e.g., directed acyclic word graphs) or binary decision diagrams (BDDs), distinct nodes may share a common subtree, preserving logical tree semantics while structurally forming a directed acyclic graph (DAG).

Take an example of a tree with an internally shared subtree. We can model the situation by letting one node own the subtree and offer it at a location \(q\), while any other node that references this subtree view it as borrowed at \(q\), analogous to the case in tree rotation (\cref{case:tree-rotation}). The entire structure is then viewed as a partial normal tree with one subtree absent.

Such partial containers cannot be expressed using magic wands. Magic wands do not model partial structures directly, but instead specify that the complete structure can be reconstructed once the missing part is supplied. In this case, a “complete” tree does not exist: the shared subtree cannot be simultaneously reclaimed by all referencing nodes while the heap segments owned by those nodes remain distinct.

\paragraph{Open questions}

Our model does not answer the following questions:
\begin{itemize}
  \item How can container-internal pointers be reasoned about in concurrent programs?
  \item How can container-internal pointers be reasoned about in general graphs with unrestricted sharing?
  \item Can our model be integrated with fractional resources~\cite{boylandCheckingInterferenceFractional2003, bornatPermissionAccountingSeparation2005} to distinguish read-only from read–write access? A preliminary formulation of the logical borrow–return rule is:
  \(\setLocalJot{0}{
  \begin{aligned}
    &(\bR{(\Offered{v}{q}{qs})}{p})^{\pi} \\
    \equiv\; &\bR{(\Borrowed{q}{qs})}{p} \star (\bR{(\Owned{v})}{q})^{\pi} \\
  \end{aligned}
  }\\ \)
  where \(\pi \in (0, 1\rbrack\) is the permission fraction.
  \item How can container invariants be systematically tracked and enforced across borrow–return transitions?
  \item How can logical models, representation predicates and container-specific transition lemmas be automatically derived for containers with borrowable subparts?  We suspect that existing work on contexts~\cite{huetZipper1997,minamideFunctionalRepresentationData1998,mcbrideDerivativeRegularType2001,abbottDerivativesContainers2003,leijenTailRecursionModulo2023} and \textit{lenses}~\cite{bohannonBoomerangResourcefulLenses2008,fosterBidirectionalProgrammingLanguages2009} could nicely combine with our approach.
\end{itemize}



\note{potential discussion: cross-container sharing}


\note{potential discussion: top-level borrow (will not include in this paper)}


\note{potential discussion: container-specific borrowing-state-transition lemmas (How the transition bubble up from the bottom to top abstraction) }

\note{potential discussion: disjoint borrows and nested borrows}

\note{potential discussion: alternative ownership-borrowing formulations (owned/missing, offered/borrowed, owned/borrowed)}




\section{Other Related Work}\label{sec:related-work}

Beyond the work discussed in \cref{sec:background}, previous work differs from ours by focusing on automation~\cite{astrauskasLeveragingRustTypes2019,costeaConciseReadonlySpecifications2020}, on read-only sharing~\cite{costeaConciseReadonlySpecifications2020}, or on modeling specific access patterns~\cite{astrauskasLeveragingRustTypes2019,hoAeneasRustVerification2022}.

The \textbf{iterated separating conjunction}~\cite{reynoldsSeparationLogicLogic2002,müllerAutomaticVerificationIterated2016} generalizes \(\star\) to range over sets of heap locations. It is particularly useful to specify unstructured containers, which it represents as a collection of elements with all internal pointers exposed.  In contrast, logical pinning supports selective exposure.

\textbf{Authoritative cameras}~\cite{jungIrisMonoidsInvariants2015,jungIrisGroundModular2018} in Iris maintain a coherent global view of a resource by designating an authoritative token that can mutate the state and multiple fragment tokens that provide consistent read-only views of its parts.\todo{comparison?}

\textbf{Rust borrows} permit either one mutable reference or multiple immutable references to each object or object field, but not both.
Standard mutable borrows freeze the entire container, so ad hoc APIs have been developed to allow simultaneous borrows (\texttt{pick2\_mut}, \texttt{get\_\-disjoint\_\-mut}~\cite{therustprojectdevelopersSliceGet_disjoint_mutRust2025}), but they either require unsafe code, or have runtime costs, or have usage restrictions (e.g., most require obtaining all references simultaneously
).

\textbf{Deferred borrows}~\cite{fallinSafeFlexibleAliasing2020} are a borrowing system that ensures the safety of a common workaround for the limitations of borrowing: using paths (indices in an array, keys of a map, \ldots) instead of pointers.  In combination with custom container types and custom path-returning APIs, deferred borrows enable safe path references to multiple elements in a container, with actual borrows delayed until access time, at which point the path is (re-)resolved into a pointer.  In contrast, logical pinning is a purely logical discipline.

In \textbf{RefinedRust}~\cite{gaherRefinedrust2024}, a value can be logically marked as borrowed by replacing it with a borrow name. In contrast, logical pinning records the location of the borrowed value.

\section{Conclusion}

\textit{Logical pinning} is a new lightweight borrowing model that makes exposed internal pointers explicit in the logical representation of containers, enabling direct reasoning about pointer validity while preserving modularity.  This technique unifies several ad hoc proof patterns and supports precise yet modular specifications for common idioms such as pointer caching, expose-then-mutate, and consecutive lookups.

Our case studies show that it simplifies existing proofs and allow new patterns to be verified, all without reformulating the underlying separation logic. We implemented the approach in CFML to demonstrate its soundness and practicality. Our mechanization is available online for reference.

\section*{Data Availability Statement}

The formalization of the logical-pinning model and associated case studies are available at \url{https://github.com/epfl-systemf/logical-pinning}. The artifact~\cite{artifact} of this paper is available at \url{https://doi.org/10.5281/zenodo.17815704}.

\appendix

\section{The Subtree-Compare Example}\label{app:subtree-compare}

Let \(f\) be the logical comparison function, and \(\AllSubtreeAvl{\bt}\) assert that every subtree of \(\bt\) is available. The specification of \texttt{compare} uses the non-separating conjunction in its precondition to handle potentially overlapping subtrees:
\begin{flalign*}
  &\AllSubtreeAvl{\bt_1} \wedge \AllSubtreeAvl{\bt_2} \rightarrow \\
  &\tstate{(\bReprTreeS{R}{\bt_1}{q_1} \star H_1) \hand (\bReprTreeS{R}{\bt_2}{q_2} \star H_2)}\; \\
  &\quad \codeAppTwo{\texttt{compare}}{q_1}{q_2}\; \\
  &\tstatex{r}{\hpure{r = \appTwo{f}{\bt_1}{\bt_2}} \star \bReprTreeS{R}{\bt_2}{q_2} \star H_2}
\end{flalign*}
\note{[space] Can save space here}

Starting from \(\bReprTreeS{R}{\bt}{p}\), the state after performing the two \texttt{lookup} operations is as follows:
\[\bReprTreeS{R}{\substb{\substb{\bt}{k_1}{\Pin{\bt[k_1]}{q_1}}}{k_2}{\Pin{\bt[k_2]}{q_2}}}{p}\]
In unification with \texttt{compare}'s precondition, \(H_1\) is the partial tree with \(\bt[k_1]\) borrowed (\(H_2\) is obtained analogously):
\[\bReprTreeS{R}{\substb{\bt}{k_1}{\BorrowedOne{q_1}}}{p}\]
A logical return step after \texttt{compare} can restore \(\bt\).

\section{Proof of Swapping Two Elements in a List}\label{app:full-swap-proof}

\newcommand{\temp}{\subst{l}{n_1}{\Offered{v_1}{q_1}{qs_1}}}

Assume \texttt{memswap} swaps the contents of the cells pointed to by two distinct pointers, or does nothing if given the same pointer. In the following, we give the correctness proof of the element-swapping program; due to space constraints, we omit the case where \(n_1 = n_2\), which follows the same proof style. In the proof, we highlight \stepborrowtext{logical borrow steps}, \stepreturntext{logical return steps} and \stepforgettext{logical forget steps} using different colors.
\def\myspace{\mspace{35mu}}
\begin{flalign*}
& n_1 \neq n_2 \wedge n_1 < \Length{l} \wedge n_2 < \Length{l} \rightarrow \\
& \Nth{n_1}{l} = \Available{v_1}{qs_1} \wedge
  \Nth{n_2}{l} = \Available{v_2}{qs_2} \rightarrow \\
&\myspace \hlstate{\tstate{\ReprCListE{l}{p}}} \\
&\myspace \texttt{\keyword{let} \(q_1\) \(\eqdef\) nth\_elem\_ptr(\(p\), \(n_1\)) \keyword{in}} \\
&\myspace \hlstate{\tstate{\ReprCListE{\temp}{p}}} \\
&\myspace \texttt{\keyword{let} \(q_2\) \(\eqdef\) nth\_elem\_ptr(\(p\), \(n_2\)) \keyword{in}} \\
&\myspace \hlstate{\begin{mstate}
        \appOne{\reprCListE}{&\temp} \\
        &\;\appOne{[n_2 := \Offered{v_2}{q_2}{qs_2}]}{p}
        \end{mstate}} \\
&\; \stepborrow{\equiv} \stepborrow{\begin{mstate}
        &\appOne{\reprCListE}{\subst{l}{n_1}{\Borrowed{q_1}{qs_1}}} \\
        &\qquad\mspace{20mu} \appOne{[n_2 := \Borrowed{q_2}{qs_2}]}{p} \\
        &\quad \star \Cell{v_1}{q_1} \star \Cell{v_2}{q_2}
        \end{mstate}} \\
&\myspace \texttt{\ \ memswap(\(q_1\), \(q_2\))} \\
&\myspace \hlstate{\begin{mstate}
        &\appOne{\reprCListE}{\subst{l}{n_1}{\Borrowed{q_1}{qs_1}}} \\
        &\qquad\mspace{20mu} \appOne{[n_2 := \Borrowed{q_2}{qs_2}]}{p} \\
        &\quad \star \Cell{v_2}{q_1} \star \Cell{v_1}{q_2}
        \end{mstate}} \\
&\; \stepreturn{\equiv} \stepreturn{\begin{mstate}
        \appOne{\reprCListE}{&\subst{l}{n_1}{\Offered{v_2}{q_1}{qs_1}}} \\
        &\; \appOne{[n_2 := \Offered{v_1}{q_2}{qs_2}]}{p} \\
        \end{mstate}}
\end{flalign*}

\section{Proof of Binary-Tree Left Rotation}\label{app:full-tree-rotation-proof}

\def\myspace{\mspace{30mu}}
\begin{flalign*}
&\myspace \hlstate{\tstate{
  \ReprTreeS{R}{(\NodeS{x}{\btreel}{(\Available{t_r}{qs_r})})}{p}
}} \\
&\myspace \texttt{\ \  \keyword{let} pr := p.rtree \keyword{in}} \\
&\myspace \hlstate{\tstate{
  \ReprTreeS{R}{(\NodeS{x}{\btreel}{(\Offered{t_r}{p_r}{qs_r})})}{p}
}} \\
&\stepborrow{\equiv} \stepborrow{\begin{mstate}
  & \ReprTreeS{R}{(\NodeS{x}{\btreel}{(\Borrowed{p_r}{qs_r})})}{p} \\
  &\quad \star \ReprTreeS{R}{t_r}{p_r}
  \end{mstate}} \\
&\myspace \texttt{\ \ \textbf{if} not (\_is\_empty pr) \textbf{then} (} \\
&\myspace \hlstate{\begin{mstate}
  & \ReprTreeS{R}{(\NodeS{x}{\btreel}{(\Borrowed{p_r}{qs_r})})}{p} \\
  &\quad \star \ReprTreeS{R}{(\NodeS{x_r}{\btreerl}{\btreerr})}{p_r}
  \end{mstate}} \\
&\myspace \texttt{\ \ \ \ \keyword{let} prl := pr.ltree \keyword{in}} \\
&\myspace \hlstate{\begin{mstate}
  & \ReprTreeS{R}{(\NodeS{x}{\btreel}{(\Borrowed{p_r}{qs_r})})}{p} \\
  &\quad \star \ReprTreeS{R}{(\NodeS{x_r}{(\Pin{\btreerl}{p_{rl}})}{\btreerr})}{p_r}
  \end{mstate}} \\
&\myspace \texttt{\ \ \ \ p.rtree <- prl} \\
&\myspace \hlstate{\begin{mstate}
  & \ReprTreeS{R}{(\NodeS{x}{\btreel}{(\BorrowedOne{p_{rl}})})}{p} \\
  &\quad \star \hexistsx{q} \bReprTreeS{R}{(\Borrowed{p_r}{qs_r})}{q}\\
  &\quad \star \ReprTreeS{R}{(\NodeS{x_r}{(\Pin{\btreerl}{p_{rl}})}{\btreerr})}{p_r}
  \end{mstate}} \\
&\mspace{12mu} \hlstate{\equiv} \mspace{6mu}
  \hlstate{\begin{mstate}
  & \ReprTreeS{R}{(\NodeS{x}{\btreel}{(\BorrowedOne{p_{rl}})})}{p} \\
  &\quad \star \hpure{\Eqlocs{q}{qs_r}{p_r}} \\
  &\quad \star \ReprTreeS{R}{(\NodeS{x_r}{(\Pin{\btreerl}{p_{rl}})}{\btreerr})}{p_r}
  \end{mstate}} \\
&\myspace \texttt{\ \ \ \ pr.ltree <- p} \\
&\myspace \hlstate{\begin{mstate}
  & \ReprTreeS{R}{(\NodeS{x}{\btreel}{(\BorrowedOne{p_{rl}})})}{p} \\
  &\quad \star \hpure{\Eqlocs{q}{qs_r}{p_r}} \\
  &\quad \star \ReprTreeS{R}{(\NodeS{x_r}{(\BorrowedOne{p})}{\btreerr})}{p_r} \\
  &\quad \star \hexistsx{q} \bReprTreeS{R}{(\Pin{\btreerl}{p_{rl}})}{q}
  \end{mstate}} \\
&\mspace{12mu} \hlstate{\vdash} \mspace{6mu}
  \hlstate{\begin{mstate}
  & \ReprTreeS{R}{(\NodeS{x}{\btreel}{(\BorrowedOne{p_{rl}})})}{p} \\
  &\quad \star \hpure{\Eqlocs{q}{qs_r}{p_r}} \\
  &\quad \star \ReprTreeS{R}{(\NodeS{x_r}{(\BorrowedOne{p})}{\btreerr})}{p_r} \\
  &\quad \star \bReprTreeS{R}{\btreerl}{p_{rl}}
\end{mstate}} \\
&\stepreturn{\equiv} \stepreturn{\begin{mstate}
  & \ReprTreeS{R}{(\NodeS{x}{\btreel}{(\Pin{\btreerl}{p_{rl}})})}{p} \\
  &\quad \star \hpure{\Eqlocs{q}{qs_r}{p_r}} \\
  &\quad \star \ReprTreeS{R}{(\NodeS{x_r}{(\BorrowedOne{p})}{\btreerr})}{p_r} \\
  \end{mstate}} \\
&\stepforget{\vdash} \stepforget{\begin{mstate}
  & \ReprTreeS{R}{(\NodeS{x}{\btreel}{\btreerl})}{p} \\
  &\quad \star \hpure{\Eqlocs{q}{qs_r}{p_r}} \\
  &\quad \star \ReprTreeS{R}{(\NodeS{x_r}{(\BorrowedOne{p})}{\btreerr})}{p_r} \\
  \end{mstate}} \\
&\stepreturn{\equiv} \stepreturn{\begin{mstate}
  & \ReprTreeS{R}{(\NodeS{x_r}{\\ &\qquad (\OfferedOne{(\NodeS{x}{\btreel}{\btreerl})}{p})}{\btreerr})}{p_r} \\
  & \quad \star \hpure{\Eqlocs{q}{qs_r}{p_r}}
  \end{mstate}} \\
&\mspace{12mu} \hlstate{\equiv} \mspace{6mu}
  \hlstate{\tstate{\bReprTreeS{R}{(\LeftRotate{(\Offered{t}{p}{\nil})})}{p_r}}} \\
&\myspace \texttt{\ \ ) \textbf{else} p} \\
&\myspace \hlstate{\begin{mstate}
  & \ReprTreeS{R}{(\NodeS{x}{\btreel}{(\Borrowed{p_r}{qs_r})})}{p} \\
  &\quad \star \ReprTreeS{R}{\leafS}{p_r}
  \end{mstate}} \\
&\mspace{3mu} \stepreturn{\equiv} \stepreturn{ \tstate{\ReprTreeS{R}{(\NodeS{x}{\btreel}{(\Offered{\leafS}{p_r}{qs_r})})}{p}}} \\
&\mspace{3mu} \stepforget{\vdash\mspace{2mu}} \stepforget{\tstate{
  \ReprTreeS{R}{(\NodeS{x}{\btreel}{(\Available{\leafS}{qs_r})})}{p}}} \\
&\mspace{12mu} \hlstate{\equiv} \mspace{6mu}
  \hlstate{\begin{mstate}
  & \bReprTreeS{R}{(\OfferedOne{\\ &\qquad (\NodeS{x}{\btreel}{(\Available{\leafS}{qs_r})})}{p})}{p}
  \end{mstate}} \\
&\mspace{12mu} \hlstate{\equiv} \mspace{6mu} \hlstate{\tstate{\bReprTreeS{R}{(\LeftRotate{(\OfferedOne{t}{p})})}{p_r}}}\end{flalign*}

\begin{acks}

We thank Jade Philipoom for early discussions on this topic. We also thank Aurèle Barrière, Arthur Charguéraud, Gregory Malecha, Alexandre Moine, François Pottier, and the anonymous reviewers for their valuable feedback and suggestions.

\end{acks}

\clearpage
\balance

\bibliographystyle{ACM-Reference-Format}
\bibliography{main-base}

@unpublished{mcbrideDerivativeRegularType2001,
  type = {Manuscript},
  title = {The Derivative of a Regular Type Is Its Type of One-Hole Contexts},
  author = {McBride, Conor},
  year = {2001},
  url = {http://strictlypositive.org/diff.pdf},
  langid = {english}
}

@misc{caoProofPearlMagic2019,
  title = {Proof {{Pearl}}: {{Magic Wand}} as {{Frame}}},
  shorttitle = {Proof {{Pearl}}},
  author = {Cao, Qinxiang and Wang, Shengyi and Hobor, Aquinas and Appel, Andrew W.},
  year = {2019},
  month = sep,
  number = {arXiv:1909.08789},
  eprint = {1909.08789},
  publisher = {arXiv},
  doi = {10.48550/arXiv.1909.08789},
  url = {http://arxiv.org/abs/1909.08789},
  urldate = {2023-02-20},
  archiveprefix = {arXiv}
}

@inproceedings{charguéraudHigherorderRepresentationPredicates2016,
  title = {Higher-Order Representation Predicates in Separation Logic},
  booktitle = {Proceedings of the 5th {{ACM SIGPLAN Conference}} on {{Certified Programs}} and {{Proofs}}},
  author = {Chargu{\'e}raud, Arthur},
  year = {2016},
  month = jan,
  series = {{{CPP}} 2016},
  pages = {3--14},
  publisher = {Association for Computing Machinery},
  address = {New York, NY, USA},
  doi = {10.1145/2854065.2854068},
  url = {https://dl.acm.org/doi/10.1145/2854065.2854068},
  urldate = {2025-08-18},
  isbn = {978-1-4503-4127-1},
  langid = {english}
}

@inproceedings{reynoldsSeparationLogicLogic2002,
  title = {Separation {{Logic}}: {{A Logic}} for {{Shared Mutable Data Structures}}},
  shorttitle = {Separation {{Logic}}},
  booktitle = {Proceedings 17th {{Annual IEEE Symposium}} on {{Logic}} in {{Computer Science}}},
  author = {Reynolds, John C.},
  year = {2002},
  month = jul,
  pages = {55--74},
  publisher = {IEEE Computer Society},
  issn = {1043-6871},
  doi = {10.1109/LICS.2002.1029817},
  url = {https://www.computer.org/csdl/proceedings-article/lics/2002/14830055/12OmNxYbSXN},
  urldate = {2025-08-18},
  isbn = {978-0-7695-1483-3},
  langid = {english}
}

@article{huetZipper1997,
  title = {The Zipper},
  author = {Huet, G{\'e}rard},
  year = {1997},
  month = sep,
  journal = {Journal of Functional Programming},
  volume = {7},
  number = {5},
  pages = {549--554},
  issn = {1469-7653, 0956-7968},
  doi = {10.1017/S0956796897002864},
  url = {https://www.cambridge.org/core/journals/journal-of-functional-programming/article/zipper/0C058890B8A9B588F26E6D68CF0CE204},
  urldate = {2023-04-03},
  langid = {english}
}

@article{jungIrisGroundModular2018,
  title = {Iris from the Ground up: {{A}} Modular Foundation for Higher-Order Concurrent Separation Logic},
  shorttitle = {Iris from the Ground Up},
  author = {Jung, Ralf and Krebbers, Robbert and Jourdan, Jacques-Henri and Bizjak, Ale{\v s} and Birkedal, Lars and Dreyer, Derek},
  year = {2018},
  month = jan,
  journal = {Journal of Functional Programming},
  volume = {28},
  pages = {e20},
  publisher = {Cambridge University Press},
  issn = {0956-7968, 1469-7653},
  doi = {10.1017/S0956796818000151},
  url = {https://www.cambridge.org/core/journals/journal-of-functional-programming/article/iris-from-the-ground-up-a-modular-foundation-for-higherorder-concurrent-separation-logic/26301B518CE2C52796BFA12B8BAB5B5F},
  urldate = {2024-01-15},
  langid = {english}
}

@inproceedings{ohearnLocalReasoningPrograms2001c,
  title = {Local {{Reasoning}} about {{Programs}} That {{Alter Data Structures}}},
  booktitle = {Proceedings of the 15th {{International Workshop}} on {{Computer Science Logic}}},
  author = {O'Hearn, Peter W. and Reynolds, John C. and Yang, Hongseok},
  year = {2001},
  month = sep,
  series = {{{CSL}} '01},
  pages = {1--19},
  publisher = {Springer-Verlag},
  address = {Berlin, Heidelberg},
  doi = {10.5555/647851.737404},
  url = {https://dl.acm.org/doi/10.5555/647851.737404},
  urldate = {2025-07-21},
  isbn = {978-3-540-42554-0}
}

@misc{SpdlogFastLogging2025,
  title = {Spdlog: {{Fast C}}++ Logging Library.},
  year = {2025},
  month = may,
  journal = {spdlog: Fast C++ logging library},
  url = {https://github.com/gabime/spdlog/tree/v1.x},
  urldate = {2025-07-23},
  lastaccessed = {2025-07-23}
}

@misc{ContainersLibraryCppreferencecom2025,
  title = {Containers Library - Cppreference.Com},
  year = {2025},
  url = {https://en.cppreference.com/w/cpp/container.html#Iterator_invalidation},
  urldate = {2025-07-23},
  lastaccessed = {2025-07-23}
}

@phdthesis{fosterBidirectionalProgrammingLanguages2009,
  title = {Bidirectional Programming Languages},
  author = {Foster, John Nathan},
  year = {2009},
  langid = {english},
  school = {University of Pennsylvania}
}

@inproceedings{minamideFunctionalRepresentationData1998,
  title = {A Functional Representation of Data Structures with a Hole},
  booktitle = {Proceedings of the 25th {{ACM SIGPLAN-SIGACT}} Symposium on {{Principles}} of Programming Languages},
  author = {Minamide, Yasuhiko},
  year = {1998},
  month = jan,
  series = {{{POPL}} '98},
  pages = {75--84},
  publisher = {Association for Computing Machinery},
  address = {New York, NY, USA},
  doi = {10.1145/268946.268953},
  url = {https://dl.acm.org/doi/10.1145/268946.268953},
  urldate = {2025-08-18},
  isbn = {978-0-89791-979-1}
}

@article{wangCertifyingGraphmanipulatingPrograms2019a,
  title = {Certifying Graph-Manipulating {{C}} Programs via Localizations within Data Structures},
  author = {Wang, Shengyi and Cao, Qinxiang and Mohan, Anshuman and Hobor, Aquinas},
  year = {2019},
  month = oct,
  journal = {Proceedings of the ACM on Programming Languages},
  volume = {3},
  number = {OOPSLA},
  pages = {171:1--171:30},
  doi = {10.1145/3360597},
  url = {https://dl.acm.org/doi/10.1145/3360597},
  urldate = {2025-08-18}
}

@inproceedings{boylandCheckingInterferenceFractional2003,
  title = {Checking Interference with Fractional Permissions},
  booktitle = {Proceedings of the 10th International Conference on {{Static}} Analysis},
  author = {Boyland, John},
  year = {2003},
  month = jun,
  series = {{{SAS}}'03},
  pages = {55--72},
  publisher = {Springer-Verlag},
  address = {Berlin, Heidelberg},
  doi = {10.5555/1760267.1760273},
  urldate = {2025-08-18},
  isbn = {978-3-540-40325-8}
}

@inproceedings{costeaConciseReadonlySpecifications2020,
  title = {Concise Read-Only Specifications for Better Synthesis of Programs with Pointers},
  booktitle = {Programming {{Languages}} and {{Systems}}},
  author = {Costea, Andreea and Zhu, Amy and Polikarpova, Nadia and Sergey, Ilya},
  editor = {M{\"u}ller, Peter},
  year = {2020},
  pages = {141--168},
  publisher = {Springer International Publishing},
  address = {Cham},
  doi = {10.1007/978-3-030-44914-8_6},
  isbn = {978-3-030-44914-8},
  langid = {english}
}

@inproceedings{bornatPermissionAccountingSeparation2005,
  title = {Permission Accounting in Separation Logic},
  booktitle = {Proceedings of the 32nd {{ACM SIGPLAN-SIGACT}} Symposium on {{Principles}} of Programming Languages},
  author = {Bornat, Richard and Calcagno, Cristiano and O'Hearn, Peter and Parkinson, Matthew},
  year = {2005},
  month = jan,
  series = {{{POPL}} '05},
  pages = {259--270},
  publisher = {Association for Computing Machinery},
  address = {New York, NY, USA},
  doi = {10.1145/1040305.1040327},
  url = {https://dl.acm.org/doi/10.1145/1040305.1040327},
  urldate = {2025-08-18},
  isbn = {978-1-58113-830-6}
}

@inproceedings{fallinSafeFlexibleAliasing2020,
  title = {Safe, {{Flexible Aliasing}} with {{Deferred Borrows}}},
  booktitle = {34th {{European Conference}} on {{Object-Oriented Programming}} ({{ECOOP}} 2020)},
  author = {Fallin, Chris},
  editor = {Hirschfeld, Robert and Pape, Tobias},
  year = {2020},
  series = {Leibniz {{International Proceedings}} in {{Informatics}} ({{LIPIcs}})},
  volume = {166},
  pages = {30:1--30:26},
  publisher = {Schloss Dagstuhl -- Leibniz-Zentrum f{\"u}r Informatik},
  address = {Dagstuhl, Germany},
  issn = {1868-8969},
  doi = {10.4230/LIPIcs.ECOOP.2020.30},
  url = {https://drops.dagstuhl.de/entities/document/10.4230/LIPIcs.ECOOP.2020.30},
  urldate = {2025-08-18},
  isbn = {978-3-95977-154-2}
}

@article{ohearnSeparationLogic2019,
  title = {Separation Logic},
  author = {O'Hearn, Peter},
  year = {2019},
  month = jan,
  journal = {Communications of the ACM},
  volume = {62},
  number = {2},
  pages = {86--95},
  issn = {0001-0782},
  doi = {10.1145/3211968},
  url = {https://dl.acm.org/doi/10.1145/3211968},
  urldate = {2025-08-18}
}

@article{leijenTailRecursionModulo2023,
  title = {Tail {{Recursion Modulo Context}}: {{An Equational Approach}}},
  shorttitle = {Tail {{Recursion Modulo Context}}},
  author = {Leijen, Daan and Lorenzen, Anton},
  year = {2023},
  month = jan,
  journal = {Proceedings of the ACM on Programming Languages},
  volume = {7},
  number = {POPL},
  pages = {40:1152--40:1181},
  doi = {10.1145/3571233},
  url = {https://dl.acm.org/doi/10.1145/3571233},
  urldate = {2025-08-18}
}

@phdthesis{charguéraudModernEyeSeparation2023,
  type = {Thesis},
  title = {A {{Modern Eye}} on {{Separation Logic}} for {{Sequential Programs}}},
  author = {Chargu{\'e}raud, Arthur},
  year = {2023},
  month = feb,
  url = {https://inria.hal.science/tel-04076725},
  urldate = {2025-08-18},
  langid = {english},
  school = {Universit{\'e} de Strasbourg}
}

@inproceedings{müllerAutomaticVerificationIterated2016,
  title = {Automatic {{Verification}} of {{Iterated Separating Conjunctions Using Symbolic Execution}}},
  booktitle = {Computer {{Aided Verification}}},
  author = {M{\"u}ller, Peter and Schwerhoff, Malte and Summers, Alexander J.},
  editor = {Chaudhuri, Swarat and Farzan, Azadeh},
  year = {2016},
  pages = {405--425},
  publisher = {Springer International Publishing},
  address = {Cham},
  doi = {10.1007/978-3-319-41528-4_22},
  isbn = {978-3-319-41528-4},
  langid = {english}
}

@inproceedings{abbottDerivativesContainers2003,
  title = {Derivatives of {{Containers}}},
  booktitle = {Typed {{Lambda Calculi}} and {{Applications}}},
  author = {Abbott, Michael and Altenkirch, Thorsten and Ghani, Neil and McBride, Conor},
  editor = {Hofmann, Martin},
  year = {2003},
  pages = {16--30},
  publisher = {Springer},
  address = {Berlin, Heidelberg},
  doi = {10.1007/3-540-44904-3_2},
  isbn = {978-3-540-44904-1},
  langid = {english}
}

@article{hoAeneasRustVerification2022,
  title = {Aeneas: {{Rust}} Verification by Functional Translation},
  shorttitle = {Aeneas},
  author = {Ho, Son and Protzenko, Jonathan},
  year = {2022},
  month = aug,
  journal = {Proceedings of the ACM on Programming Languages},
  volume = {6},
  number = {ICFP},
  pages = {116:711--116:741},
  doi = {10.1145/3547647},
  url = {https://dl.acm.org/doi/10.1145/3547647},
  urldate = {2025-08-18}
}

@inproceedings{bohannonBoomerangResourcefulLenses2008,
  title = {Boomerang: Resourceful Lenses for String Data},
  shorttitle = {Boomerang},
  booktitle = {Proceedings of the 35th Annual {{ACM SIGPLAN-SIGACT}} Symposium on {{Principles}} of Programming Languages},
  author = {Bohannon, Aaron and Foster, J. Nathan and Pierce, Benjamin C. and Pilkiewicz, Alexandre and Schmitt, Alan},
  year = {2008},
  month = jan,
  series = {{{POPL}} '08},
  pages = {407--419},
  publisher = {Association for Computing Machinery},
  address = {New York, NY, USA},
  doi = {10.1145/1328438.1328487},
  url = {https://dl.acm.org/doi/10.1145/1328438.1328487},
  urldate = {2025-09-12},
  isbn = {978-1-59593-689-9}
}

@misc{therustprojectdevelopersSplittingBorrowsRustonomicon2025,
  title = {Splitting {{Borrows}} - {{The Rustonomicon}}},
  author = {{The Rust Project Developers}},
  year = {2025},
  url = {https://doc.rust-lang.org/nomicon/borrow-splitting.html},
  urldate = {2025-09-12},
  lastaccessed = {2025-09-12}
}

@misc{therustprojectdevelopersSliceGet_disjoint_mutRust2025,
  title = {Slice::Get\_disjoint\_mut - {{The Rust Standard Library}}},
  author = {{The Rust Project Developers}},
  year = {2025},
  url = {https://doc.rust-lang.org/std/primitive.slice.html#method.get_disjoint_mut},
  urldate = {2025-09-12},
  lastaccessed = {2025-09-12}
}

@inproceedings{jungIrisMonoidsInvariants2015,
  title = {Iris: {{Monoids}} and {{Invariants}} as an {{Orthogonal Basis}} for {{Concurrent Reasoning}}},
  shorttitle = {Iris},
  booktitle = {Proceedings of the 42nd {{Annual ACM SIGPLAN-SIGACT Symposium}} on {{Principles}} of {{Programming Languages}}},
  author = {Jung, Ralf and Swasey, David and Sieczkowski, Filip and Svendsen, Kasper and Turon, Aaron and Birkedal, Lars and Dreyer, Derek},
  year = {2015},
  month = jan,
  series = {{{POPL}} '15},
  pages = {637--650},
  publisher = {Association for Computing Machinery},
  address = {New York, NY, USA},
  doi = {10.1145/2676726.2676980},
  url = {https://dl.acm.org/doi/10.1145/2676726.2676980},
  urldate = {2025-09-12},
  isbn = {978-1-4503-3300-9}
}

@inproceedings{astrauskasLeveragingRustTypes2019,
  title = {Leveraging Rust Types for Modular Specification and Verification},
  booktitle = {Object-{{Oriented Programming Systems}}, {{Languages}}, and {{Applications}} ({{OOPSLA}})},
  author = {Astrauskas, Vytautas and M{\"u}ller, Peter and Poli, Federico and Summers, Alexander J.},
  year = {2019},
  month = oct,
  volume = {3},
  pages = {147:1--147:30},
  publisher = {ACM},
  doi = {10.1145/3360573},
  url = {https://dl.acm.org/doi/10.1145/3360573},
  urldate = {2025-09-12}
}

@article{gaherRefinedrust2024,
	title = {{RefinedRust}: {A} {Type} {System} for {High}-{Assurance} {Verification} of {Rust} {Programs}},
	volume = {8},
	shorttitle = {{RefinedRust}},
	url = {https://dl.acm.org/doi/10.1145/3656422},
	doi = {10.1145/3656422},
	number = {PLDI},
	urldate = {2025-11-18},
	journal = {Proceedings of the ACM on Programming Languages},
	author = {Gäher, Lennard and Sammler, Michael and Jung, Ralf and Krebbers, Robbert and Dreyer, Derek},
	month = jun,
	year = {2024},
	pages = {192:1115--192:1139},
}

@software{artifact,
  author       = {Guan, Yawen and Pit-Claudel, Clément},
  title        = {Artifact: Precise Reasoning about Container-Internal Pointers with Logical Pinning},
  month        = dec,
  year         = 2025,
  publisher    = {Zenodo},
  version      = {1.0},
  doi          = {10.5281/zenodo.17815704},
  url          = {https://doi.org/10.5281/zenodo.17815704},
}

\setlength{\jot}{\oldJot}

\end{document}